\documentclass[useAMS,usenatbib,twocolumn,fleqn]{mnras}
\usepackage[T1]{fontenc}
\usepackage{ae,aecompl}
\usepackage{natbib, graphics, epsfig}
\usepackage{stmaryrd}

\usepackage{times}
\usepackage{amsmath}
\usepackage{amssymb}
\usepackage{textcomp}
\usepackage{verbatim}

\newcommand{\hmsun}{\,h^{-1}\,{\rm M_\odot}}
\newcommand{\msun}{{\,\rm M_\odot}}

\newcommand{\kms}{\,{\rm km}\,{\rm s}^{-1}}

\newcommand{\Gyr}{\,{\rm Gyr}}

\newcommand{\Mpc}{\,{\rm Mpc}}

\title[Intra-cluster metals in IllustrisTNG] 
{The uniformity and time-invariance of the intra-cluster metal distribution in galaxy clusters from the IllustrisTNG simulations}

\author[M.~Vogelsberger et al.]
{Mark Vogelsberger$^1$\thanks{Alfred P. Sloan Fellow, email:mvogelsb@mit.edu}, 
Federico Marinacci$^1$, \!\!
Paul Torrey$^1$\thanks{Hubble Fellow},
Shy Genel$^{2,3}$,
Volker Springel$^{4,5}$,
\newauthor
Rainer Weinberger$^{4}$,
R\"udiger Pakmor$^{4}$,
Lars Hernquist$^{6}$,
Jill Naiman$^{6}$,
Annalisa Pillepich$^{7}$,
\newauthor
Dylan Nelson$^{8}$
\vspace*{0.2cm}\\
$^1${Department of Physics, Kavli Institute for Astrophysics and Space Research, Massachusetts Institute of Technology, Cambridge, MA 02139, USA}\\
$^2${Center for Computational Astrophysics, Flatiron Institute, 162 Fifth Avenue, New York, NY 10010, USA}\\
$^3${Columbia Astrophysics Laboratory, Columbia University, 550 West 120th Street, New York, NY 10027, USA}\\
$^4${Heidelberg Institute for Theoretical Studies, Schloss-Wolfsbrunnenweg 35, D-69118 Heidelberg, Germany}\\
$^5${Zentrum f{\"u}r Astronomie der Universit{\"a}t Heidelberg, ARI, M{\"o}nchhofstr. 12-14, D-69120 Heidelberg, Germany}\\
$^6${Harvard--Smithsonian Center for Astrophysics, 60 Garden Street, Cambridge, MA 02138}\\
$^7${Max-Planck-Institut f{\"u}r Astronomie, K{\"o}nigstuhl 17, 69117 Heidelberg, Germany}\\
$^8${Max-Planck-Institut f{\"u}r Astrophysik, Karl-Schwarzschild-Str. 1, 85741 Garching, Germany}\\}

\begin{document}

\pagerange{\pageref{firstpage}--\pageref{lastpage}}
\pubyear{2017}

\maketitle

\label{firstpage}

\begin{abstract}
The distribution of metals in the intra-cluster medium encodes important
information about the enrichment history and formation of galaxy clusters. Here
we explore the metal content of clusters in IllustrisTNG -- a new suite of
galaxy formation simulations building on the Illustris project. Our cluster
sample contains $20$ objects in TNG100 -- a $\sim (100\,{\rm Mpc})^3$ volume
simulation with $2 \times 1820^3$ resolution elements, and $370$ objects in
TNG300 -- a $\sim (300\,{\rm Mpc})^3$ volume simulation with $2 \times 2500^3$
resolution elements. The $z=0$ metallicity profiles agree with observations,
and the enrichment history is consistent with observational data going beyond
$z \sim 1$, showing nearly no metallicity evolution. The abundance profiles
vary only minimally within the cluster samples, especially in the outskirts
with a relative scatter of $\sim 15\%$. The average metallicity profile
flattens towards the center, where we find a logarithmic slope of $-0.1$
compared to $-0.5$ in the outskirts. Cool core clusters have more centrally
peaked metallicity profiles ($\sim 0.8$ solar) compared to non-cool core
systems ($\sim 0.5$ solar), similar to observational trends. Si/Fe and O/Fe
radial profiles follow positive gradients. The outer abundance
profiles do not evolve below $z \sim 2$, whereas the inner profiles flatten
towards $z=0$. More than $\sim 80\%$ of the metals in the intra-cluster medium
have been accreted from the proto-cluster environment, which has been enriched
to $\sim 0.1$ solar already at $z\sim 2$. We conclude that the intra-cluster
metal distribution is uniform among our cluster sample, nearly time-invariant
in the outskirts for more than $10\,{\rm Gyr}$, and forms through a universal
enrichment history.
\end{abstract}


\section{Introduction}\label{sec:intro}

Galaxy clusters are the largest collapsed objects in the Universe and are made up of hundreds
to thousands of galaxies, hot plasma ($T \sim 10^8{\rm K}$), and dark matter.
There is about a factor of ten more mass in the intra-cluster medium (ICM) than
in stars~\citep[e.g.,][]{Gonzalez2013}, and $90\%$ of the total mass is in dark
matter (DM)~\citep[e.g.,][]{Allen2008,Chiu2016}.  Due to its high temperature
the ICM emits X-ray radiation allowing spectral X-ray observations of galaxy
clusters to provide an immense amount of insight into the temperature and
density structure of the hot gas within the ICM~\citep[for a review,
see][]{Bohringer2010}.  Furthermore, X-ray line observations provide information on
the abundance and distribution of metals within the ICM.

First observations by ASCA and BeppoSAX found that the average metallicity of
the ICM is around $\sim 1/3$ in solar units~\citep[e.g.,][]{DeGrandi2001}.
Since the ICM dominates the baryonic mass budget of clusters there is more mass
in metals in the ICM than in all cluster galaxies
combined~\citep[][]{Renzini2014}.  These first results were followed by much
deeper and better resolved observations with XMM-Newton, Chandra, Suzaku and
Hitomi, which allowed a characterisation of the spatial distribution of metals
and their evolution in much greater detail~\citep[e.g.,][]{Werner2006,
Sanderson2009, Matsushita2011, Bulbul2012, Molendi2016, McDonald2016,
Mantz2017}. 

Those studies have revealed a striking universality and constancy of the
metallicity and elemental abundance distributions within the ICM.  Except for
the innermost parts of the ICM most clusters have similar metallicity profiles
with shallow negative gradients. The detailed metallicity profile near the
cluster center depends on the thermodynamic profile. For cool core clusters the
metallicity peaks at the center whereas non-cool core clusters exhibit shallow
gradients all the way towards the center without such a
peak~\citep[e.g.,][]{DeGrandi2001, DeGrandi2004, Baldi2007, Leccardi2008,
Johnson2011, Elkholy2015}. Consequently, the level of core entropy
anti-correlates with the central enrichment level~\citep[][]{Leccardi2010}.
The central excess in cool core clusters could, for example, have been produced
by the stellar population of the brightest cluster
galaxy~\citep[][]{Boehringer2004}.  Most observations also find that there is
only a weak dependence of the average ICM metallicity on the cluster
temperature; e.g.~\cite{Baldi2012} find $\langle Z \rangle \propto \langle k T \rangle^{0.06 \pm 0.16}$; i.e.
the average metallicity does not depend strongly on cluster mass or X-ray
luminosity although~\cite{Yates2017} recently found a slight anti-correlation
for the temperature-iron abundance relation. 

More recent observations also quantified the cluster enrichment history to
higher redshifts finding that the median average metallicity since $z \sim 1$
changes by less then $40\%$ compared to its present-day
value~\citep[e.g.,][]{Baldi2012, Ettori2015, McDonald2016, Mantz2017}; i.e.  there is no
strong evolutionary trend visible in metallicity going back in time.
Observations of the outskirts of galaxy clusters~\citep[see][for a review]{Reprich2013} also revealed remarkable
uniformity with azimuth~\citep[][]{Werner2013, Simionescu2015, Urban2017} supporting the
idea that those regions should have been enriched early on.

\begin{table*}
\begin{tabular}{llccrrrccc}
\hline
{\bf IllustrisTNG Simulation} & Run &   \multicolumn{2}{c}{box side length}
& $N_{\rm gas}$  & $N_{\rm dm}$ & $N_{\rm tracer}$
  & $m_{\rm b}$ & $m_{\rm dm}$ & $\epsilon$ \\
&        &   $[h^{-1}{\rm Mpc}]$  & $[{\rm Mpc}]$  & & &
 & $[h^{-1}{\rm M}_\odot]$ & $[h^{-1}{\rm M}_\odot]$
&  $[h^{-1}{\rm kpc}]$ \\
\hline
\hline
{\bf TNG300} &  TNG300-1  & 205  & 302.6 &
 $2500^3$ & $2500^3$ & $2 \times 2500^3$
& $7.44\times 10^6$ & $3.98\times 10^7$ & 1.0 \\
                     &  TNG300-2  & 205  & 302.6 &
 $1250^3$ & $1250^3$ & $2 \times 1250^3$
& $5.95\times 10^7$ & $3.19\times 10^8$ & 2.0 \\
                &  TNG300-3  & 205  & 302.6 &
 $625^3$ & $625^3$ & $2 \times 625^3$
& $4.76\times 10^8$ & $2.55\times 10^9$ & 4.0 \\
\hline
{\bf TNG100} &  TNG100-1  & 75  & 110.7 &
 $1820^3$ & $1820^3$ & $2 \times 1820^3$
& $9.44\times 10^5$ & $5.06\times 10^6$ & 0.5 \\
                     &  TNG100-2  & 75  & 110.7 &
 $910^3$ & $910^3$ & $2 \times 910^3$
& $7.55\times 10^6$ & $4.04\times 10^7$ & 1.0 \\
                &  TNG100-3  & 75  & 110.7 &
 $455^3$ & $455^3$ & $2 \times 455^3$
& $6.04\times 10^7$ & $3.24\times 10^8$ & 2.0 \\
\hline
\end{tabular}
\caption{The IllustrisTNG simulation suite contains three major simulations covering three different volumes,
roughly $\sim 50^3, 100^3, 300^3\,{\rm Mpc}^3$: {\bf TNG50} ($N_{\rm gas} + N_{\rm DM} = 2 \times 2160^3$,
$m_{b}=5.74\times 10^4\hmsun$), {\bf TNG100} ($N_{\rm gas} + N_{\rm DM} = 2 \times 1820^3$, $m_{\rm b} = 9.44 \times
10^5\hmsun$), and {\bf TNG300} ($N_{\rm gas} + N_{\rm DM} = 2 \times 2500^3$, $m_{b}= 7.44 \times 10^6\hmsun$). The table presents the basic numerical parameters of the IllustrisTNG simulations studied in this paper (TNG100 and TNG300): simulation volume, number of gas cells ($N_{\rm gas}$), number of dark matter particles ($N_{\rm DM}$), number of tracer particles ($N_{\rm tracer}$), baryon mass resolution ($m_{\rm b}$), dark matter mass resolution ($m_{\rm DM}$), and Plummer-equivalent gravitational softening length ($\epsilon$). For each volume, we have run different numerical resolutions spaced
by a factor of eight in mass resolution. The gravitational softening lengths $\epsilon$ refer to the
maximum physical softening length of dark matter and star particles. The
softening of gaseous cells is tied to their radius and allowed to fall below
this value. 
\label{tab:tabsims}} 
\end{table*}

The iron abundance
within the ICM can be constrained rather well through Fe-K and Fe-L emission
measurements.  However, quantifying the abundance of elements other than iron
is more challenging and more uncertain.  The most recent studies tend to find
typically rather flat X/Fe ratios~\citep[][]{Simionescu2015, Mernier2017,
Ezer2017, Simionescu2017} pointing towards a similar distribution of iron and
lighter elements in the ICM supporting the idea of an early ICM enrichment.

Observations of the detailed distribution of metals in the ICM provide
important information to help understand the metal production and cluster formation
processes. Core-collapse supernovae (SNcc) are mainly responsible for the
production of light elements such as oxygen, neon, magnesium and silicon. SNIa,
on the other hand, produce large amounts of heavier elements like iron and
nickel~\citep[for a review, see][]{Nomoto2013}.  These two distinct production
channels furthermore operate on different timescales in releasing the metals.
Observations of relative abundance ratios can therefore help to constrain the
various enrichment processes in the ICM.  However, clusters provide a rather
complex astrophysical and dynamical environment, where lots of different
physical processes like  galactic winds, active galactic nuclei (AGN) feedback,
and gas stripping lead to mixing and a redistribution of metals in the
ICM~\citep[][]{Churazov2001, Rebusco2005, Simionescu2008, Simionescu2009,
Kirkpatrick2015}.  Despite this complexity, a key interpretation of the
uniformity of the iron distribution and the flat abundance ratio profiles
within the ICM is that early enrichment ($z>2$) might be required to achieve
this level of homogeneity throughout the ICM. Otherwise, more recent
astrophysical events within the ICM should cause more significant metal
abundance variations. Indeed, some clusters show a metallicity distribution
indicative of such recent events with fluctuations in the abundance patterns.
For example, \cite{Kirkpatrick2015} found higher metallicities along the axes
of X-ray cavities, which could be caused by AGN feedback.  

Some simple theoretical attempts of chemical and population synthesis modelling
failed to simultaneously match the constraints on ICM metallicity and on the
stellar mass to light ratios~\citep[][]{Renzini2014}. These authors claim that
the iron yield in the largest clusters ($M_{\rm 500,crit} > 10^{14}\msun$) must
be a few times the solar value. Other studies came to similar
conclusions~\citep[][]{Nagashima2005, Arrigoni2010}. Various solutions to this
modelling problem have been suggested; e.g., changes to the shape of the
initial stellar mass function, increasing the efficiency of iron production in
SNIa, or more efficient metal ejection from galaxies.  Hydrodynamical
simulations on the other hand provide a more self-consistent and more complete
approach for studying the growth and enrichment of galaxy clusters. Over the
last decade those simulations had have mixed success in explaining some of the
observed metallicity relations and abundance patterns~\citep[][]{Tornatore2007,
Cora2008, Rasia2008, Fabjan2010, McCarthy2010, Planelles2014, Rasia2015,
Martizzi2016, Biffi2017}.  \cite{Martizzi2016}, for example, find too low
metallicities in the ICM by up to factors of about $\sim 2 - 5$ depending on
cluster-centric distance.  \cite{Rasia2015} and \cite{Biffi2017}, on the other
hand, find reasonable agreement with the observed iron and metallicity
profiles. However, they have to resort to artificial thermal conduction, and
this heat transport can flatten the metal distribution as~\cite{Kannan2017}
recently pointed out.  It is therefore important to gauge the effects of
thermal conduction properly before drawing conclusions. Other simulations, for
example those presented in \cite{Tornatore2007}, correctly capture aspects of the ICM
enrichment, but do not convincingly demonstrate that they also predict the correct stellar masses of galaxies within
clusters; i.e.  they do not provide a consistent picture. Besides the actual
galaxy formation model, results might also depend on the numerical scheme and
resolution as discussed by \cite{Martizzi2016}.  So far no hydrodynamical
simulation could simultaneously achieve the following goals:  produce a
reasonable galaxy population from dwarf to cluster scales (e.g., correct
stellar masses, galaxy colors, clustering); produce the right amount of ICM
enrichment (e.g., correct metallicity gradients and overall enrichment in
clusters, abundance ratios, cool core/non-cool core dichotomy, enrichment history); and demonstrate
the latter two points over a large unbiased sample of numerically well-resolved galaxy
clusters. 

\begin{figure*}
\centering
\includegraphics[width=1.0\textwidth]{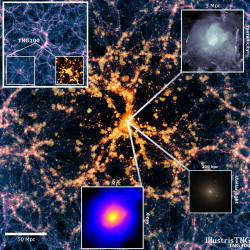}\\
\vspace{0.1cm}
\hspace{0.0075cm}
\includegraphics[width=0.334\textwidth]{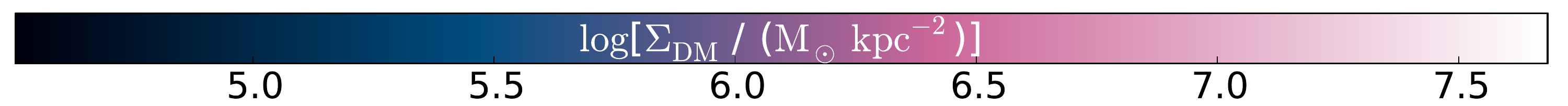}
\hspace{-0.2cm}
\includegraphics[width=0.334\textwidth]{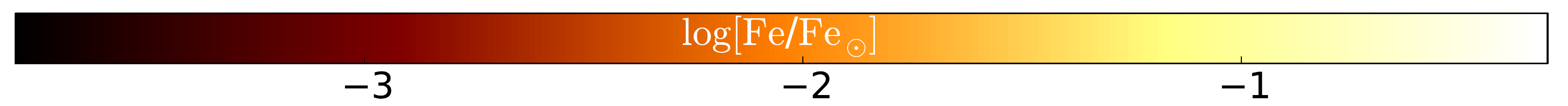}
\hspace{-0.2cm}
\includegraphics[width=0.334\textwidth]{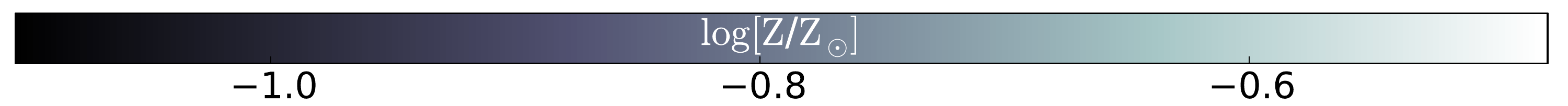}
\caption{Overview of two of the three IllustrisTNG simulations. The main panel shows a slice of {\bf TNG300} with a thickness of $22.13\Mpc$ and an extent of the
full TNG300 box ($302.63\Mpc$) at $z=0$. We show the DM density field (outer part) and
iron distribution (inner part). The maps are centered on the most massive
cluster of TNG300 ($M_{\rm 200,crit}=1.54\times 10^{15}\msun$). The iron distribution
traces the large scale matter distribution with gas outflows also enriching the
intergalactic medium. The upper right inset shows the total metal distribution in a $3\Mpc$ region around the most massive cluster in TNG300. The lower right inset
shows the stellar light of the central galaxy of the corresponding cluster. The bottom inset shows the X-ray emission of the hot gas around this most massive cluster. In the upper left panel
we present the DM density fields of {\bf TNG100} and TNG300 in relative scale, and the iron map of TNG100.}
\vspace{-0.5cm}
\label{fig:fullbox}
\end{figure*}
\begin{figure*}
\centering
\includegraphics[width=0.24\textwidth]{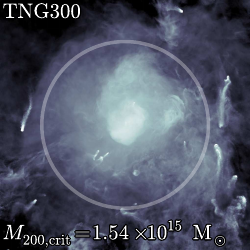}
\includegraphics[width=0.24\textwidth]{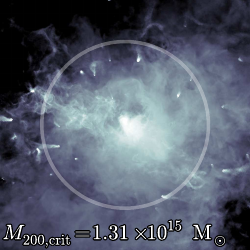}
\includegraphics[width=0.24\textwidth]{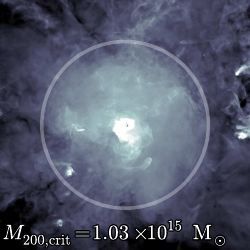}
\includegraphics[width=0.24\textwidth]{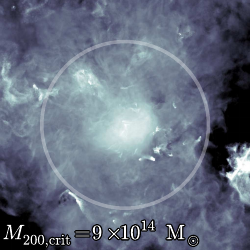}
\includegraphics[width=0.24\textwidth]{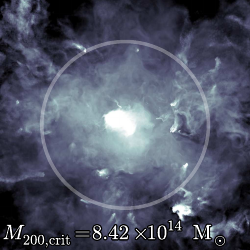}
\includegraphics[width=0.24\textwidth]{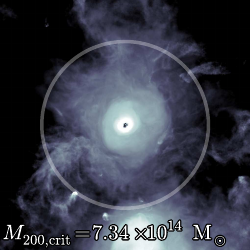}
\includegraphics[width=0.24\textwidth]{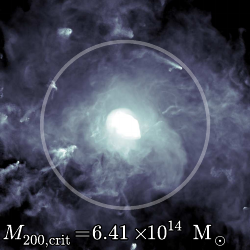}
\includegraphics[width=0.24\textwidth]{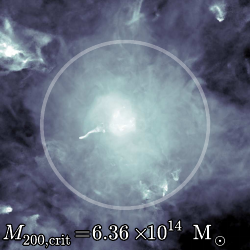}
\includegraphics[width=0.24\textwidth]{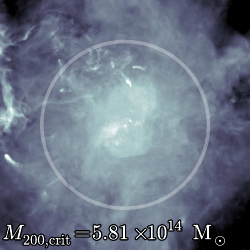}
\includegraphics[width=0.24\textwidth]{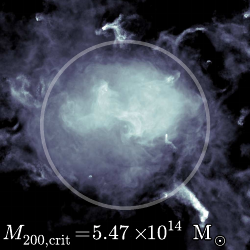}
\includegraphics[width=0.24\textwidth]{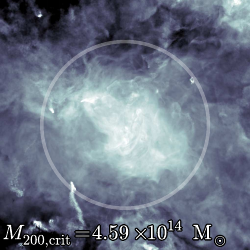}
\includegraphics[width=0.24\textwidth]{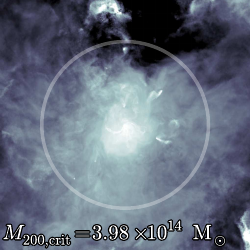}\\
\vspace{0.25cm}
\includegraphics[width=0.24\textwidth]{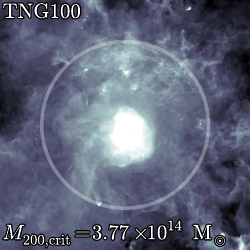}
\includegraphics[width=0.24\textwidth]{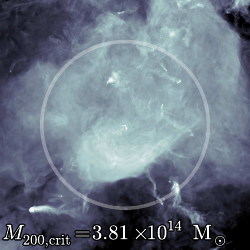}
\includegraphics[width=0.24\textwidth]{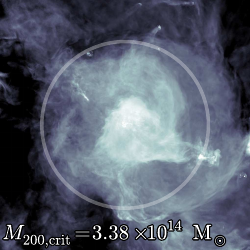}
\includegraphics[width=0.24\textwidth]{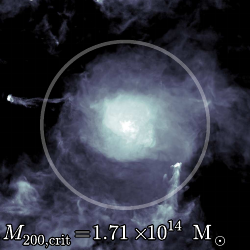}
\includegraphics[width=0.24\textwidth]{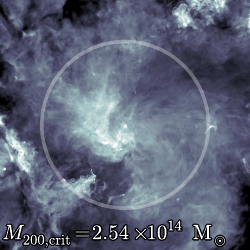}
\includegraphics[width=0.24\textwidth]{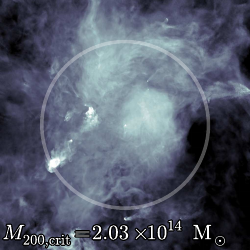}
\includegraphics[width=0.24\textwidth]{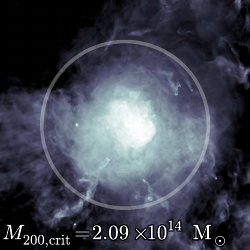}
\includegraphics[width=0.24\textwidth]{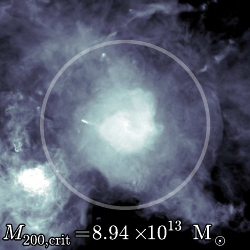}
\includegraphics[width=0.8\textwidth]{figures/maps/colorbar_fof_0_12_z0_Z_TNG300.pdf}
\caption{Metallicity maps for twelve (eight) massive clusters of TNG300 (TNG100) at $z=0$. The
extent and depth of each map is $3 \times r_{\rm 500,crit}$. Circles denote
$r_{\rm 500, crit}$. The virial mass ($M_{\rm 200,crit}$) is indicated in each
panel. The average metallicity within all clusters is approximately constant,
but the metal distribution shows local variations among the different
clusters. In all clusters the metallicity is increasing towards the center and
reaches values around $\sim 0.1\,{\rm Z}_\odot$ at the virial radius. In some
clusters stream like features are also visible due to metal stripping from
infalling cluster galaxies.}
\vspace{-0.5cm}
\label{fig:metal_maps}
\end{figure*}

The goal of this paper is to fill this gap and present a first assessment of
the metal content of the ICM of galaxy clusters in IllustrisTNG -- an ongoing follow-up simulation campaign of the Illustris project containing
various large-scale hydrodynamical galaxy formation simulations. IllustrisTNG
allows us to study the chemical composition of the ICM of a large sample of
galaxy clusters, which is much larger than any
previously studied cluster sample at that numerical resolution.  Our study is therefore distinctly different from
previous theoretical attempts to study the ICM enrichment for several reasons: (i) We employ a
state-of-the-art galaxy formation model that passes a large number of
observational tests from dwarf to cluster scales.  (ii) The numerical
resolution of our cluster samples is higher than in most of the previous
studies. (iii) The cluster sample size of our simulation is one to two orders
of magnitude larger than any previous attempt studying the ICM enrichment.
(iv) Our cluster sample is large enough to probe significantly different
assembly and merging histories thereby minimising potential biases. (v) Our simulation code has been demonstrated to be efficient
and accurate in solving the hydrodynamical equations.

The paper is structured as follows. In Section~\ref{sec:sims} we briefly
describe our methods and the simulation suite that we are using for our
analysis. In Section~\ref{sec:present_icm} we present a detailed analysis of
the present-day metallicity distribution in the ICM of our clusters. We inspect the
redshift evolution of the metal content in Section~\ref{sec:past_icm}, where we also analyse the origin of the metals in the ICM. Finally,
we present our conclusions in Section~\ref{sec:conclusions}.

\section{Simulations}\label{sec:sims}

\begin{figure}
\centering
\includegraphics[width=0.235\textwidth]{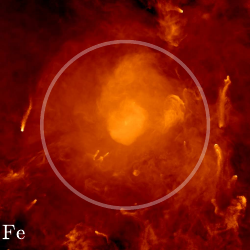}
\includegraphics[width=0.235\textwidth]{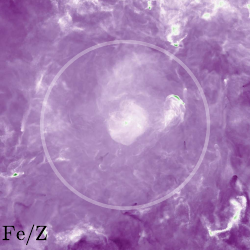}\\
\includegraphics[width=0.235\textwidth]{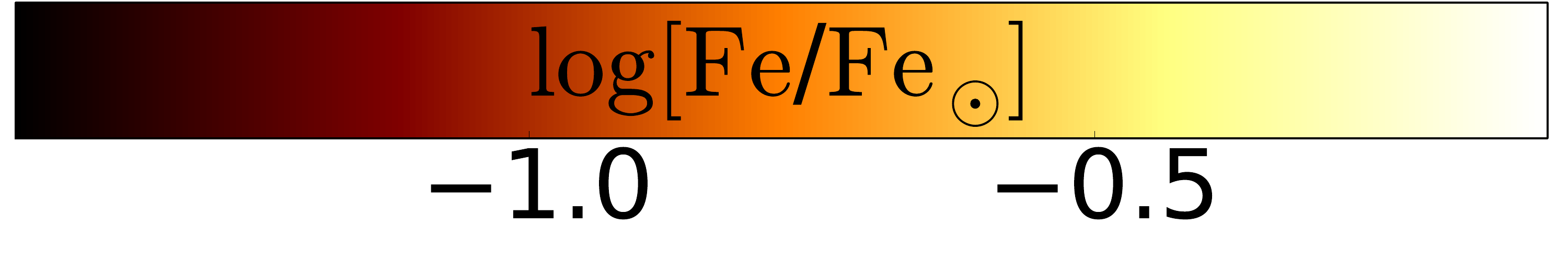}
\includegraphics[width=0.235\textwidth]{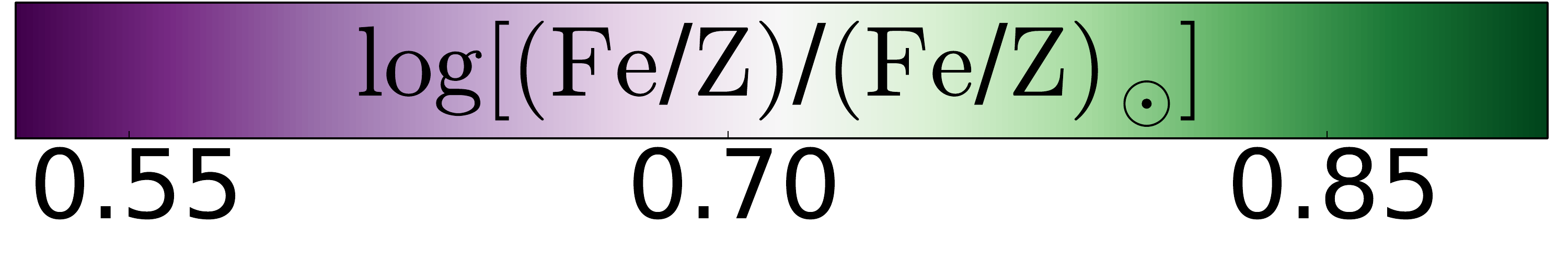}\\
\includegraphics[width=0.235\textwidth]{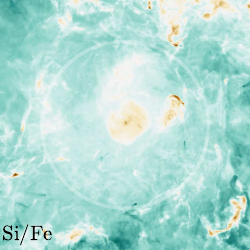}
\includegraphics[width=0.235\textwidth]{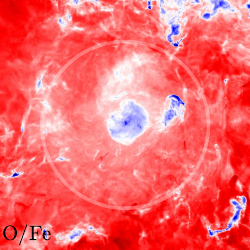}\\
\includegraphics[width=0.235\textwidth]{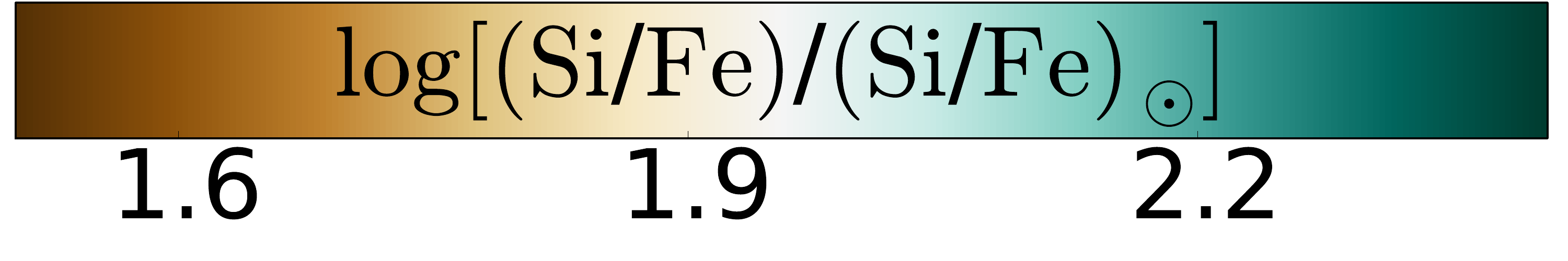}
\includegraphics[width=0.235\textwidth]{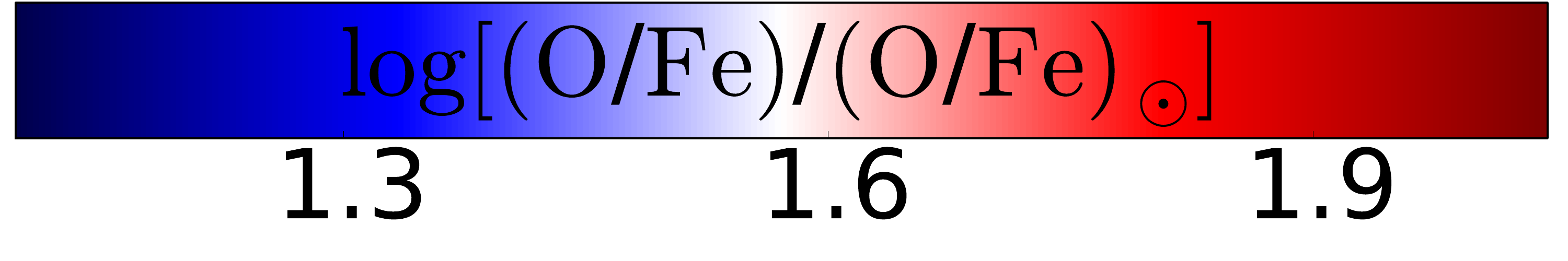}
\caption{Maps for iron (upper left), iron over all metals
(upper right), silicon over iron (lower left), and oxygen over iron (lower
right) for the most massive cluster of TNG300 at $z=0$. The extent and depth of
each map is $3 \times r_{\rm 500,crit}$. Circles denote $r_{\rm 500, crit}$.  The
element ratios are spatially not uniform, but show small-scale variations. The
details of those patterns are caused by the different enrichment channels of
the various elements. Interestingly, some detailed features of the ratio distributions are
quite pronounced, but less visible in the iron distribution.}
\vspace{-0.4cm}
\label{fig:element_maps}
\end{figure}

In the following we study the most massive halos of the IllustrisTNG
simulations~\citep[][]{Springel2017, Marinacci2017, Naiman2017, Pillepich2017a, Nelson2017}. IllustrisTNG is the follow-up project of
Illustris~\citep{Vogelsberger2014, Illustris, Genel2014, Sijacki2015}. The Illustris data has been made publicly available~\citep[][]{Nelson2015a}.  The IllustrisTNG
simulation suite consists of a set of cosmological simulations with the
following cosmological parameters that have been chosen in accordance with Planck constraints~\citep[][]{Planck2016}: $\Omega_{\rm m} = 0.3089$, $\Omega_{\rm b} =
0.0486$, $\Omega_{\Lambda} = 0.6911$, $H_0 = 100\,h\,\kms \Mpc^{-1} =
67.74\,\kms \Mpc^{-1}$, $\sigma_{8} = 0.8159$, and $n_{\rm s} = 0.9667$.

The IllustrisTNG simulation suite contains three major simulations covering three different periodic, uniformly sample volumes,
roughly ${\sim 50^3}, 100^3, 300^3\,{\rm Mpc}^3$: {\bf TNG50} ($N_{\rm gas} + N_{\rm DM} = 2 \times 2160^3$,
$m_{b}=5.74\times 10^4\hmsun$), {\bf TNG100} ($N_{\rm gas} + N_{\rm DM} = 2 \times 1820^3$, $m_{\rm b} = 9.44 \times
10^5\hmsun$), and {\bf TNG300} ($N_{\rm gas} + N_{\rm DM} = 2 \times 2500^3$, $m_{b}= 7.44 \times 10^6\hmsun$) -- the maximum resolution is indicated in parenthesis showing number of
resolution elements and baryon mass resolution.
We summarise the
numerical parameters of the TNG100 and TNG300 runs, the simulations used in this paper, in Table~\ref{tab:tabsims}.

The simulations were carried out with the moving-mesh code {\sc Arepo}
\citep{Arepo}.  The IllustrisTNG simulations employ a comprehensive module for
galaxy formation physics, which is an updated version of the Illustris
model~\citep{Vogelsberger2013, Torrey2013, Vogelsberger2014, Illustris,
Genel2014}. Details of the updated model are described in
\cite{Weinberger2017a} and \cite{Pillepich2017}. The new model features a new
radio mode AGN feedback scheme~\citep[][]{Weinberger2017a}, numerical
improvements for the convergence properties~\citep[][]{Pakmor2016}, the
inclusion of ideal magnetohydrodynamics~\citep[][]{Pakmor2013}, a retuned SN
wind model, and refinements/extensions to the chemical evolution
scheme~\citep[][]{Pillepich2017}.

The main differences between the IllustrisTNG and Illustris galaxy formation models
affecting the metal production and distribution
are: the AGN feedback implementation, the galactic
wind scalings, and the details of the stellar evolution and chemical enrichment model.
IllustrisTNG employs a kinetic AGN feedback model in the low-accretion state,
which produces black hole-driven winds, whereas
Illustris models the low-accretion state with a radio
bubble scheme~\citep[][]{Sijacki2007, Vogelsberger2013}. 
The SN-driven galactic wind characteristics have also been modified for IllustrisTNG:
wind material has now a thermal component; it is injected isotropically
instead of bipolar; injection occurs with a gas-metallicity dependent wind energy; winds have a
minimum wind velocity and are injected with a redshift dependent wind velocity
implying that the wind velocity and the growth of the virial halo mass have the
same scaling with redshift.  
The stellar evolution and chemical enrichment schemes also differ in a few
ways. Both models employ a \cite{Chabrier2003} IMF, but the IllustrisTNG model
assumes that stars pass through an AGB phase in the mass range $1-8\msun$,
while stars with masses between $8$ and $100\msun$ end as SNcc. The transition
in the Illustris model was $6\msun$. 
Furthermore the yield tables have also been updated going from the Illustris to the
IllustrisTNG model.  The AGB yields are extended from their original Illustris
$1-6\msun$ mass range to $1-7.5\msun$. For masses between $1-6\msun$
IllustrisTNG uses the \cite{Karakas2010} yields. For masses of $7$ and
$7.5\msun$ IllustrisTNG employs the \cite{Doherty2014} and \cite{Fishlock2014}
yields.  The largest update is to the SNII tables, which for IllustrisTNG
extend from $8$ to $120\msun$: for $13-40~\msun$ the yields are taken from~
\cite{Kobayashi2006}, which are then complemented to lower and higher masses
with the yields presented in~\cite{Portinari1998}. IllustrisTNG also employs
the \cite{Nomoto1997} yields for SNIa, which lead only to minor
differences with respect to the SNIa yields of the Illustris model. However, the IllustrisTNG SNIa rates also differ from the Illustris model based
on the changes proposed in~\cite{Marinacci2014}. The combined changes
in the stellar evolution model and yields can cause quite significant changes
in the integrated metal production. For example, the amount of produced iron over one
Hubble time differs by a factor of $\sim 2$.

In the following we will present our metallicity results in units of the solar
abundances of~\cite{Anders1989}, which is the most commonly used system of
units for presenting observational metallicity and abundance measurements.
Specifically, we assume ${\rm Z}_\odot=0.0194$, ${\rm Fe}_\odot=1.853 \times
10^{-3}$, ${\rm O}_\odot=9.593\times 10^{-3}$, ${\rm Si}_\odot=6.998 \times
10^{-4}$ gas mass fractions when presenting metallicities, iron, oxygen, and
silicon abundances. In the following we calculate based on the simulation
data the total metal mass of a cell or the mass of a given element in that cell and then divide this by the total gas mass in a cell to derive a mass fraction. The resulting mass fractions
are the divided by the corresponding solar mass fractions. We note that more recent measurements find lower solar
metallicities; e.g.  ${\rm Z}_\odot=0.0133$~\citep[][]{Lodders2003}, ${\rm
Z}_\odot=0.0141$~\citep[][]{Lodders2009}, and ${\rm
Z}_\odot=0.0134$~\citep[][]{Asplund2009}. These solar values can be used to easily convert our results into other systems of units.

\section{Present-day ICM Composition}\label{sec:present_icm}

\subsection{Metallicity and abundance ratio maps}

Before describing the metal content of individual clusters, we show in
Fig.~\ref{fig:fullbox} the large-scale distribution of DM and iron of TNG300.
The iron abundance is presented in solar units as described above.  This map
has a side-length of $302.63\Mpc$ and a thickness of $22.13\Mpc$. The central
part shows the iron distribution, whereas the outer parts visualise the DM mass
distribution. The map is focused on the most massive cluster of TNG300 ($M_{\rm
200,crit}=1.54\times 10^{15}\msun$). As expected the overall iron abundance
follows the large-scale structure. We note that the extent of the iron
distribution differs between Illustris and IllustrisTNG due to the different
AGN feedback; i.e. the new kinetic AGN feedback of IllustrisTNG does not
transport metals out as far as the Illustris bubble radio mode~\citep{Weinberger2017a}. Fig.~\ref{fig:fullbox} also contains four 
insets showing the metal distribution, X-ray emission and stellar light associated with the most
massive cluster at the center of the projection. The fourth, upper left inset compares the size of the TNG300 volume with TNG100, for which
we also show the iron distribution.
 
We begin our detailed analysis of the ICM metal enrichment and chemical
composition by visually inspecting a few massive clusters of TNG300 and TNG100.  In
Fig.~\ref{fig:metal_maps} we present metallicity maps for twelve (eight) massive clusters
of TNG300 (TNG100) at $z=0$.  These clusters span a mass range from $M_{\rm 500,
crit}=4.17 \times 10^{14} \msun$ to $M_{\rm 500, crit}=1.16 \times 10^{15}
\msun$. The corresponding radii cover a range from $r_{\rm 500, crit}=1\Mpc$ to
$r_{\rm 500, crit}=1.63\Mpc$. The circles in the different panels of
Fig.~\ref{fig:metal_maps} denote $r_{\rm 500,crit}$, and the virial masses
($M_{\rm 200,crit}$) of the clusters are indicated in each panel. The total
horizontal and vertical extent of the projection region is $3 \times r_{\rm
500, crit}$. The maps in Fig.~\ref{fig:metal_maps} demonstrate that the metal
distribution in the ICM is not fully homogeneous. However, the average
metallicity roughly approaches the typical values of about $\sim 0.1-0.5$ solar,
in agreement with observational findings. One can also see how the metallicity
increases towards the centers of clusters. 

Despite strong observational evidence for this nearly
universal metallicity of the ICM, not every simulation has been able to
actually reproduce those values and the observed shallow metallicity gradients
in the ICM. For example~\cite{Martizzi2016} find significantly too low
metallicities in their simulations. We note however that the actual amount of
metals in the ICM and the overall amplitude of the corresponding profiles
depend on the employed stellar yields that enter the chemical evolution model.
Uncertainties in these yields directly propagate into uncertainties in the
amplitude of ICM metal profiles.  One should therefore not over-interpret
the normalisation of the profiles given the large yield uncertainties.

Inspecting the maps of the different clusters in Fig.~\ref{fig:metal_maps} also
reveals that the average metallicity seems to be rather similar across this
small cluster sample. No cluster has a particularly low or high average ICM
metallicity compared to the other clusters in the sample.  We will quantify
this similarity in more detail below.  Again, this finding is in good agreement
with observational data, which indicates that, for example, the $\langle k_B T
\rangle - \langle Z \rangle$ relation of clusters has a rather shallow slope
with little scatter; i.e. there is only a weak dependence of the average
cluster metal content on its mass, temperature or X-ray luminosity. For
example, \cite{Baldi2012} find a slope of ($0.06 \pm 0.16$) for the mean trend
of the temperature-metallicity scaling relation. Despite the uniformity and
similarity of the different maps, one can also appreciate that detailed
features in each map differ from cluster to cluster. These variations are
driven by the specific formation history of the cluster as well as internal processes
like AGN feedback, SN winds, stripping processes, and self-enrichment within
the cluster environment.  Hints of these irregularities are also seen in some
recent cluster observations~\citep[][]{Kirkpatrick2015, Vagshette2016}. 

Such small-scale features and fluctuations should also be present in the
abundance patterns of other elements in the ICM. Our chemical evolution model
individually tracks a variety of chemical elements (H, He, C, N, O, Ne, Mg, Si,
Fe) beyond pure metallicity.  Maps of some of these elements are presented in
Fig.~\ref{fig:element_maps}, where we show four different views of the most
massive cluster of TNG300 with $M_{\rm 200,crit}=1.54\times 10^{15}\msun$. The
maps present from upper left to lower right: the iron (Fe) abundance, the iron
over total metallicity (Fe/Z) ratio,  the silicon over iron (Si/Fe) ratio, and
the oxygen over iron (O/Fe) ratio as indicated. These maps can directly be compared to the first
map of Fig.~\ref{fig:metal_maps}, which shows the corresponding distribution of
all metals.  Any visible differences in these maps must be attributed to the
different production and distribution mechanisms for iron, oxygen, silicon
compared to the general metal distribution mapped by Z. Silicon and oxygen have
different production channels compared to iron, with their main contribution
coming from SNcc, whereas iron is mostly produced through SNIa. SNIa produce a
large amount of iron and nickel, while SNcc are the main contributors of oxygen
and silicon. Lighter elements (carbon, nitrogen) are also produced by low- and
intermediate-mass stars during their asymptotic giant branch (AGB) phase.  Despite these differences the
maps show overall rather similar structures as the metallicity maps in
Fig.~\ref{fig:metal_maps}.  The abundance of the elements is also quite close
to solar. However, the ratio maps demonstrate that the actual abundance ratios
of elements can fluctuate within the ICM.  For example, the silicon over iron
and oxygen over iron ratios are smaller at the cluster center compared to its
outskirts at $r_{\rm 500,crit}$.  Furthermore, silicon over iron and
oxygen over iron ratios are low at the centers of strongly emitting
regions. On average the Si/Fe values are larger than O/Fe. The fact that the
Si/Fe and O/Fe ratios slightly decrease towards the center can be interpreted
as a slight decrease in enrichment due to SNcc compared to SNIa for smaller radii. We will show
below that this is indeed the case for our cluster samples in the IllustrisTNG
simulations.  

\begin{figure}
\centering
\hspace{-0.25cm}\includegraphics[width=0.49\textwidth]{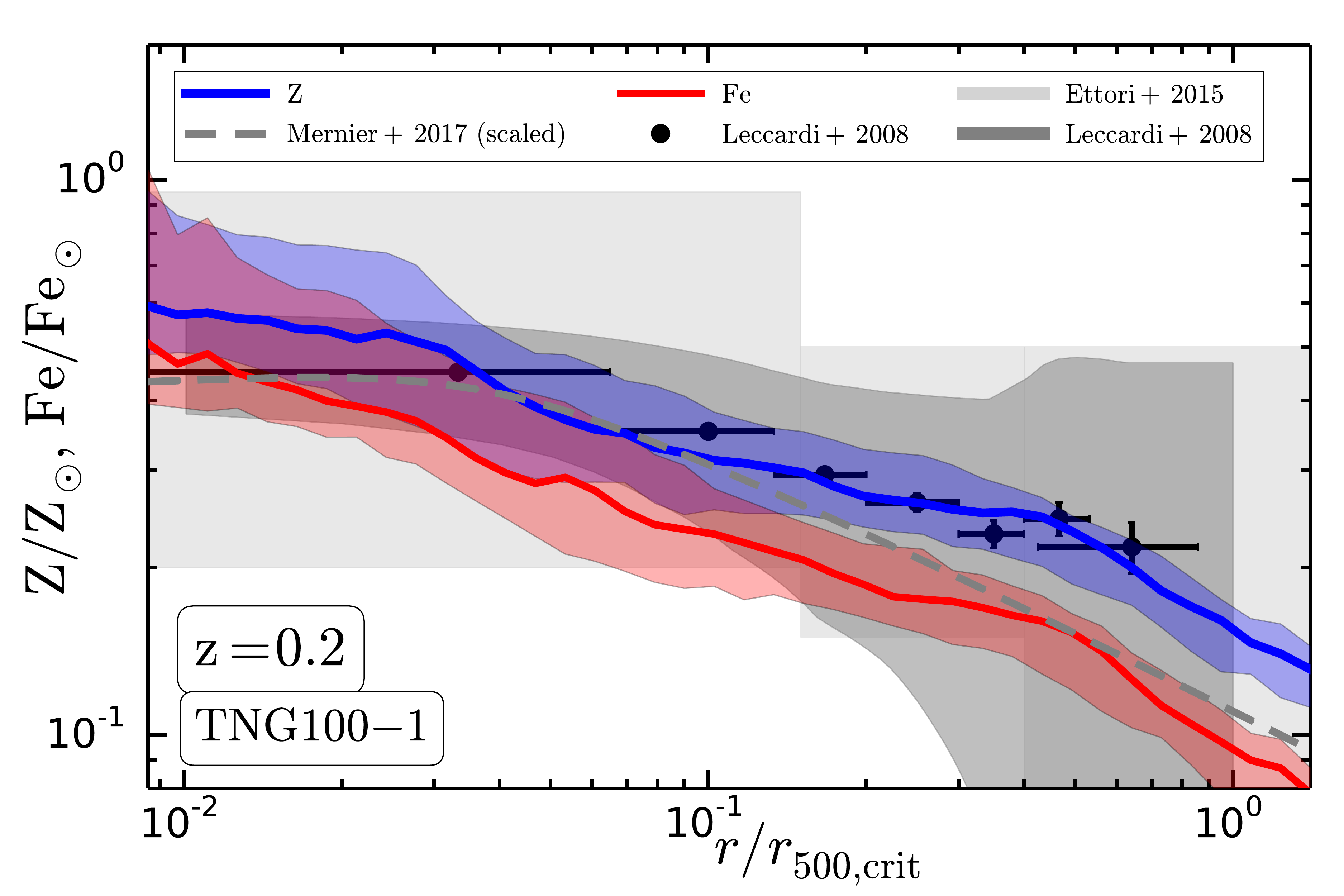}\\
\hspace{-0.25cm}\includegraphics[width=0.49\textwidth]{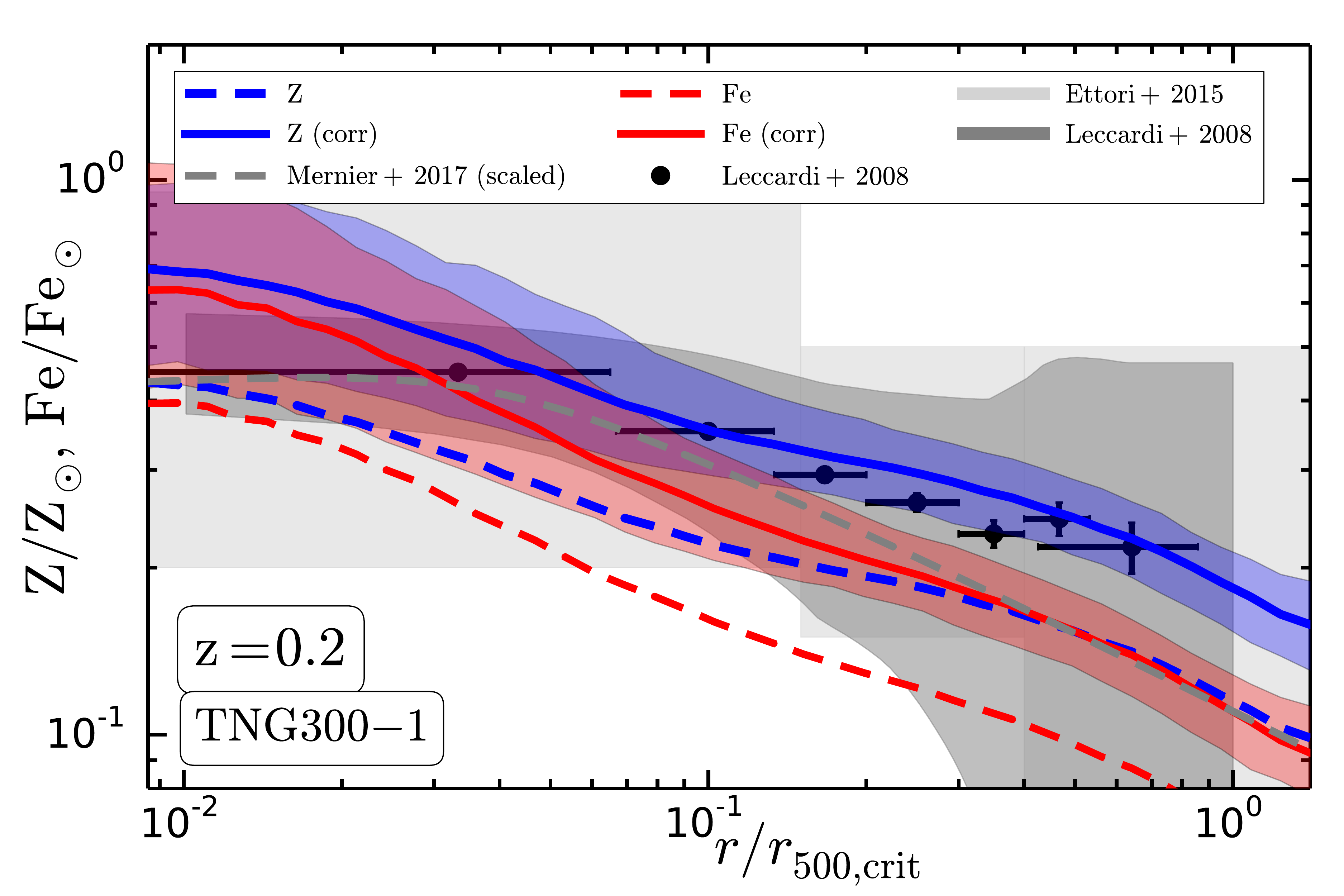}\\
\hspace{-0.25cm}\includegraphics[width=0.49\textwidth]{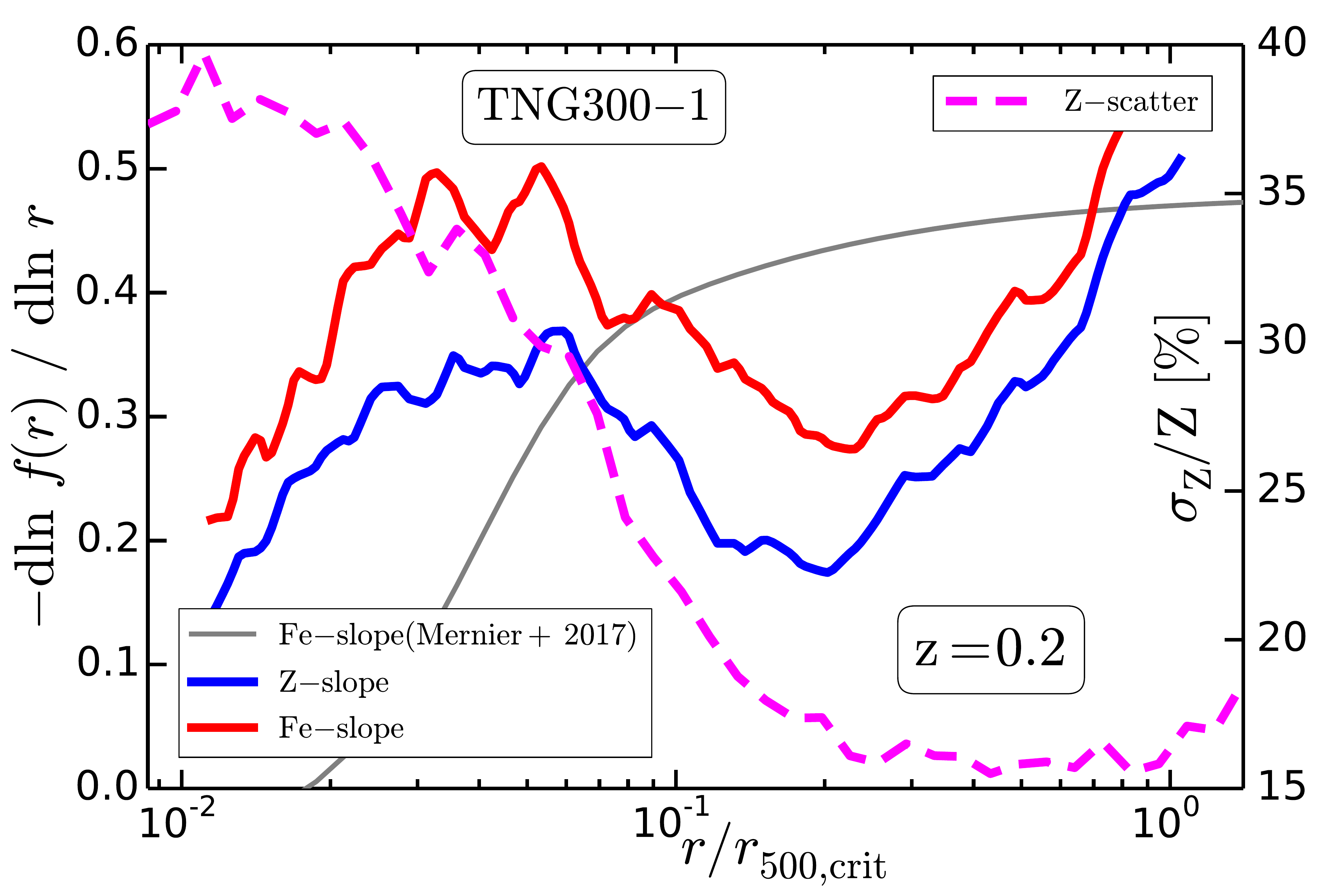}
\caption{Radial metallicity and iron profiles at $z=0.2$ for TNG100-1 (top) and
TNG300-1 (middle). The lines show median value of the mass-weighted profiles
of the cluster samples ($M_{\rm 500, crit}>10^{13.75}\msun$). This sample contains $20$ clusters for TNG100 and $370$ clusters
for TNG300.  For TNG300-1 we also present the resolution corrected (corr)
profile. The shaded regions denote the $1\sigma$
spread among the different cluster profiles of the samples. Observational data
bands are taken from \protect\cite{Leccardi2008} and \protect\cite{Ettori2015}; data points are mean values
taken from~\protect\cite{Leccardi2008}. The TNG100 profiles agree with the
observational data within the error bars. At all radii, especially in the outer
parts, the scatter between the cluster sample is small demonstrating
that the metallicity profiles are rather uniform across the full cluster
sample. The grey dashed line shows a scaled version of the empirical iron profile fit in~\protect\cite{Mernier2017}.
The bottom panel shows the negative logarithmic slopes of the
metallicity and iron profiles (solid lines), and the relative $1\sigma$
metallicity scatter as a function of radius (dashed line). The grey line shows the slope of
the empirical iron profile fit in~\protect\cite{Mernier2017}.}
\vspace{-0.5cm}
\label{fig:metal_profiles}
\end{figure}

\subsection{Radial metallicity profiles}

Observations with XMM-Newton, Chandra and Suzaku have established a variety of
insights into the metal content and metal distribution within galaxy clusters.
A general result is that radial profiles of the iron abundance show shallow
negative gradients, steeper for relaxed cool core clusters, with central
metallicity values approaching nearly solar abundances and with a global
average enrichment at a level of about $\sim 1/3$ of the solar value. It is
therefore interesting to compute metallicity and iron profiles of our cluster
samples of TNG300 and TNG100 and contrast our simulation predictions with these
observational findings.  

We present those profiles along with observational data in
Fig.~\ref{fig:metal_profiles}. For definiteness, we present here and in the following mass-weighted three-dimensional averages
of non-star forming gas in the clusters. We note that we explicitly tested that those profiles are very similar to projected emission-weighted profiles~\citep[see also][]{Biffi2017}, and that any differences are significantly smaller than uncertainties introduced due to our choice of elemental yields. Observationally CCD-resolution X-ray spectra
from Chandra and XMM-Newton put constraints on the ICM metal abundance
profiles, which are inferred from equivalent width measurement of the Fe
K$\alpha$ emission line at $6.7\,{\rm keV}$. In Fig.~\ref{fig:metal_profiles}
we show both the metallicity and iron profile since the latter is
observationally the more fundamental quantity. The lines in
Fig.~\ref{fig:metal_profiles} show the median over all mass-weighted cluster
profiles for all clusters with $M_{\rm 500,crit}>10^{13.75}\msun$. In TNG100 our cluster sample contains $20$
clusters, and in TNG300 we find $370$ clusters fulfilling the $M_{\rm 500,
crit}$ cut in cluster mass. TNG300 offers significantly more statistics due to the increased
simulation volume, which is about $20$ times larger than the TNG100 volume. The top
panel of Fig.~\ref{fig:metal_profiles} presents the results for the small
cluster sample of TNG100, while the middle panel shows the significantly larger
cluster sample of TNG300. We note that we apply the same cluster mass cut for both
simulations implying that the mean mass of the cluster samples of TNG100 and TNG300 differ due to the different simulation volumes. The shaded colored regions indicate in each panel the $1\sigma$
spread between the different individual cluster profiles by calculating the
$16\%$ and $84\%$ percentiles. All quoted $1\sigma$ ranges in the following will be computed using this percentile range.

For the comparison to observations we take data points from
the~\cite{Leccardi2008} XMM-Newton sample of $50$ objects, which contains a mixture of cool core
and non-cool core clusters with redshifts between $0.1$ and $0.4$ and average
temperatures $k T> 3.3\,{\rm keV}$. We compare this observational sample to our
simulation data at $z=0.2$, which is the mean redshift of the observational
sample. We convert the $r_{\rm 180,crit}$ data points of~\cite{Leccardi2008} to
$r_{\rm 500,crit}$ radii by assuming $r_{\rm 500,crit} \simeq 0.6 \times r_{\rm
180,crit}$, which on average holds for our cluster sample.  Besides this
observational sample we also compare to more recent observational data
from~\cite{Ettori2015}.  They provide a combined analysis of the metal content
of $83$ objects in the redshift range $0.09 - 1.39$ and with a gas temperature between $2\,{\rm keV}$ and $12.8\,{\rm keV}$.
Their data is
spatially-resolved in three radial bins with $(0 - 0.15, 0.15 - 0.4, > 0.4)\times
r_{\rm 500,crit}$, and obtained with a similar analysis technique using XMM-Newton observational data
as in~\cite{Leccardi2008} and~\cite{Baldi2012}. For the \cite{Ettori2015} data
we take the range of metallicity values for $z<0.2$ in each radial bin. The
grey shaded region in Fig.~\ref{fig:metal_profiles} shows the expected range of
metallicities for clusters within that redshift range. 

The metallicity profiles of TNG100-1 agree with the observational data
of~\cite{Leccardi2008} within the observational error bars and also lie well within the metallicity ranges
of~\cite{Ettori2015}. We note that IllustrisTNG (TNG100-1) correctly predicts the
amplitude and overall shape of the observed metallicity profiles.  More
specifically, the metallicity and iron profiles presented in
Fig.~\ref{fig:metal_profiles} typically slowly rise towards the center of the
clusters; i.e. they show a negative radial gradient following the trends also
seen in observations. At the smallest radii they reach values around $\sim
0.6$ solar in agreement with observational results.  At a radial distance of
about $r_{\rm 500, crit}$ we find average metallicities of around $\sim 0.2$
solar for the TNG100 sample and $\sim 0.1$ for the TNG300 sample.  
We note that the simulated profiles are rather steep beyond $r_{\rm 500, crit}$, which is not fully consistent with observations. We will show below
that the detailed metal distribution is affected by the details of the galaxy formation model even at these radii.
The difference in normalisation between the TNG100-1 and TNG300-1 profiles is due
to numerical resolution; i.e. the numerical resolution of TNG300 is not
sufficient to achieve metallicity profile convergence at the level of TNG100-1.
Based on convergence studies, presented below, we expect that the TNG300-1
profiles should be shifted towards larger metallicities by a factor of $1.6$ to
match the numerical resolution of TNG100-1. The corrected profiles are shown as
solid lines in the middle panel of Fig.~\ref{fig:metal_profiles}, and the
uncorrected profiles as dashed lines.  We will in the following correct the
TNG300-1 results by this factor to extrapolate to TNG100-1 resolution. The
bottom panel of Fig.~\ref{fig:metal_profiles} shows the logarithmic slopes of
the metallicity and iron profile (solid lines).  The inner region shows a
flattening of the iron profile with a slope approaching $\sim -0.2$ at a
percent of $r_{\rm 500, crit}$, whereas the profiles steepen in the outer parts
reaching slopes of $\sim -0.5$ around $\sim r_{\rm 500, crit}$. Interestingly, 
the iron profile is at all radii steeper than the total metallicity profile. We will find a
similar result below when inspecting the radial profiles of silicon and oxygen. Those are also shallower
than the iron profile, and contribute to the shallower behaviour of the overall metallicity profile. The gray line
represents the iron slope derived from the empirical iron profile fits to stacked iron measurements
presented in~\cite{Mernier2017}
based on measurements of $44$ nearby cool core galaxy clusters, groups, and ellipticals taken from the CHEERS catalogue. We also
include a scaled version of this fit, with functional form ${\rm Fe}(r) = A(r-B)^C -D \exp[-(r-E)^2/F]$, in the upper panels of Fig.~\ref{fig:metal_profiles}.
The general trend of this logarithmic slope profile is similar to the simulation data,
i.e. it also shows a flattening towards the center. However, the detailed slope
profile is rather different since it is monotonically flattening towards the
center, which is not the case for the simulation data. This can also be seen from the
upper panels, which demonstrate that the functional form of the fit does not describe the simulation
data adequately. We note however, that
the  sample size, selection and mass range of the observational sample is
rather different from the simulation sample. Specifically,
the~\cite{Mernier2017} sample contains also group scale systems and focuses on
cool core systems, which can potentially explain the discrepancies.  

\begin{figure}
\centering
\hspace{-0.25cm}\includegraphics[width=0.49\textwidth]{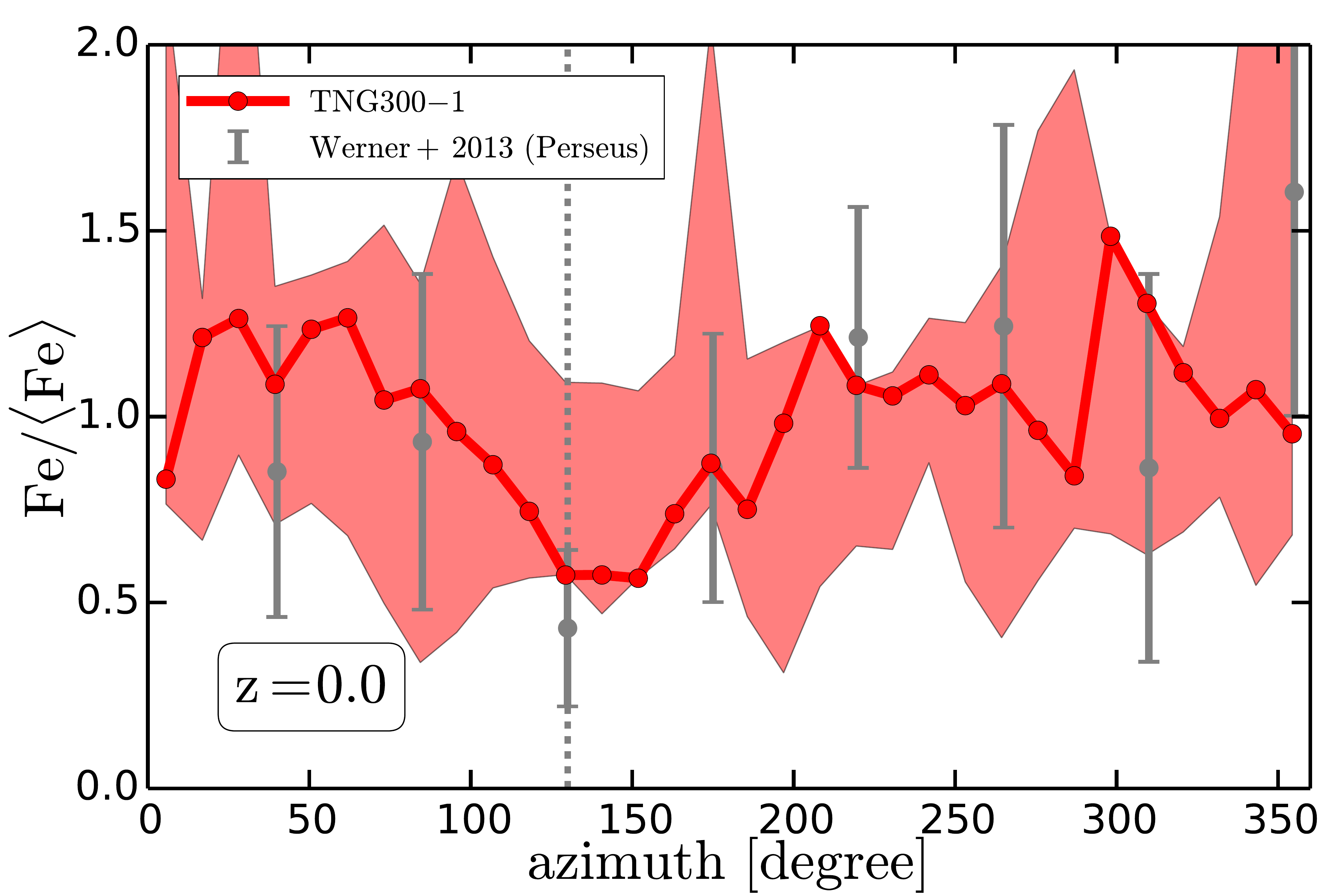}
\caption{Azimuthal normalised iron distribution for the most massive cluster in
TNG300 at $z=0$ (solid red line). The azimuthal distribution is measured at $1.4 \times r_{\rm 500, crit}$.
Observational data points are taken from eight azimuthal directions of the
Perseus cluster using Suzaku observations~\protect\citep[][]{Werner2013}. Observational error bars show $68\%$
confidence levels. The overall fluctuation level of the iron abundance
agrees between the simulation prediction and observations. For the comparison
we have identified the north direction of the Perseus observational data with
$130$ degree indicated by the vertical dashed line. The shaded regions shows the envelope of all profiles, not only $1\sigma$, for the ten most massive clusters, which all show a similar level
of fluctuations at that radial distance.}
\vspace{-0.5cm}
\label{fig:azimuth}
\end{figure}

Besides the general slope of the profiles, it is also interesting to study the
profile scatter among the different individual clusters. Here we find that
especially in the outskirts of the clusters, at radii around $\sim r_{\rm 500,
crit}$, this scatter is very small demonstrating that the iron and metallicity
profiles do not vary strongly between different clusters. The scatter is
quantified in more detail in the lower panel of Fig.~\ref{fig:metal_profiles},
where the dashed line presents the relative $1\sigma$ scatter of the
metallicity profiles as a function of radius. As described above, this scatter
is rather small in the outer parts ($\sigma_{\rm Z}/{\rm Z} \sim 15\%$), and increases to nearly $\sigma_{\rm Z}/{\rm Z} \sim 40\%$ at
one percent of $r_{\rm 500,crit}$.  We stress that, for TNG300, this scatter
includes $370$ different cluster profiles covering a wide mass range; i.e.
there is indeed strong metal uniformity in the metal profiles of clusters. In fact, over most
of the plotted radial range we find $\sigma_{\rm Z}/{\rm Z} \lesssim 35\%$. Therefore, this points towards a rather universal metallicity profile, especially in the outer
parts of the ICM. The scatter in the inner parts is larger, which is in
agreement with observations that typically show a larger degree of core
metallicity dispersion among different clusters, but uniformity in the
outskirts~\citep[][]{Elkholy2015, Mernier2016, Mernier2017}.  Also individual
cluster profiles show only small metallicity variations in the outskirts. We
demonstrate this in Fig.~\ref{fig:azimuth}, where we present the azimuthal
normalised iron distribution for the most massive cluster of TNG300 measured at
$1.4 \times r_{\rm 500, crit}$.  Observational data points are taken from eight
azimuthal directions of the Perseus cluster based on $84$ Suzaku
observations~\protect\citep{Werner2013}. The overall fluctuation strength of
the iron abundance agrees between the simulation prediction and observations.
For the comparison we have identified the north direction of the Perseus
observational data with $130$ degree as indicated by the vertical dashed line.
The shaded regions shows the envelope of all profiles for the ten most massive
clusters, which all show a similar level of fluctuations at that radial
distance.  Both observations and our simulation consistently find a rather low fluctuation
level in the outskirts of galaxy clusters.
Typically, this is interpreted as a common origin of the metals in the
outskirts due to some early enrichment within the proto-cluster environment. We
will get back to this point below.

\begin{figure}
\centering
\hspace{-0.25cm}\includegraphics[width=0.49\textwidth]{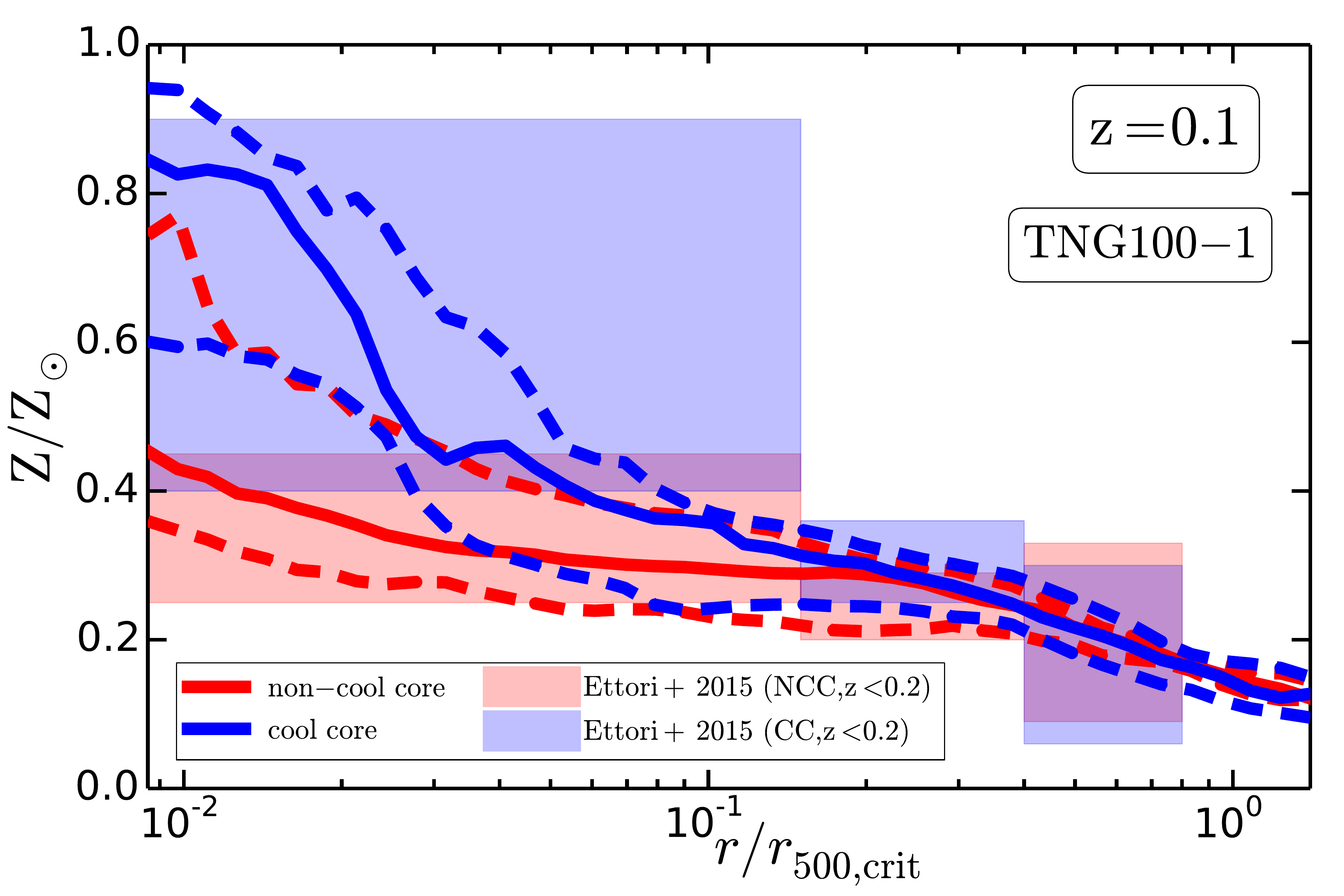}
\caption{Median metallicity profiles for TNG100-1 divided into cool core and
non-cool core clusters at $z=0.1$,  where we require a cool core cluster
to have a central entropy of less than $30\,{\rm keV}\,{\rm cm}^{-2}$. Observational data bands are taken from
\protect\cite{Ettori2015}. The median cool core and non-cool core metallicity profiles of the
simulation are in broad agreement with the observational results and show a
similar trend; i.e. cool core clusters show rising metallicity profiles towards
the center. In agreement with observations the profiles agree beyond
$0.15 \times r_{\rm 500,crit}$. Towards the center the cool core profiles approach
metallicities nearly twice as large as those of the non-cool core systems. Dashed lines show the $1\sigma$ scatter for both the cool core
and non-cool core cluster samples.}
\vspace{-0.5cm}
\label{fig:cool_core}
\end{figure}

\begin{figure*}
\centering
\includegraphics[width=0.498\textwidth]{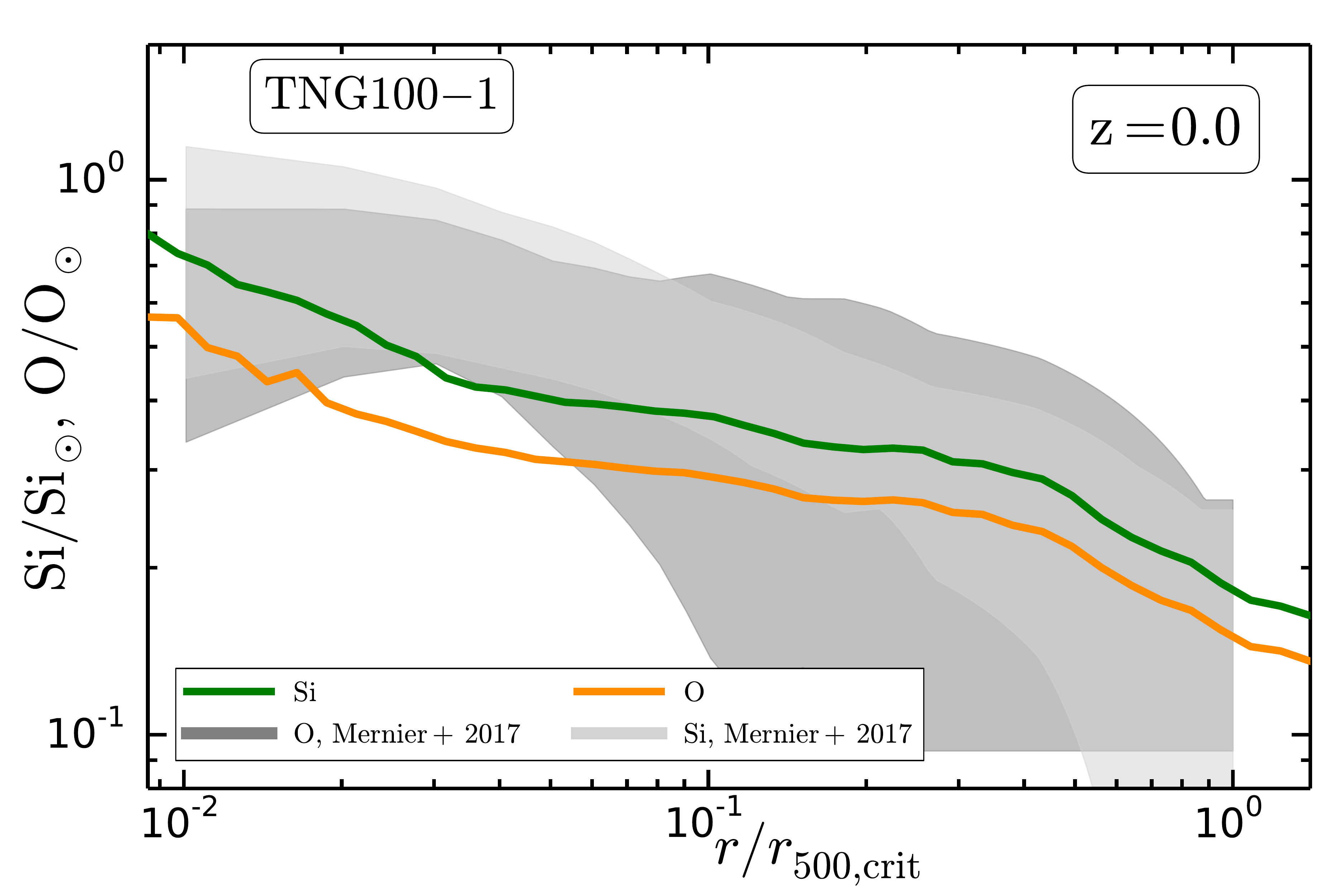}
\includegraphics[width=0.498\textwidth]{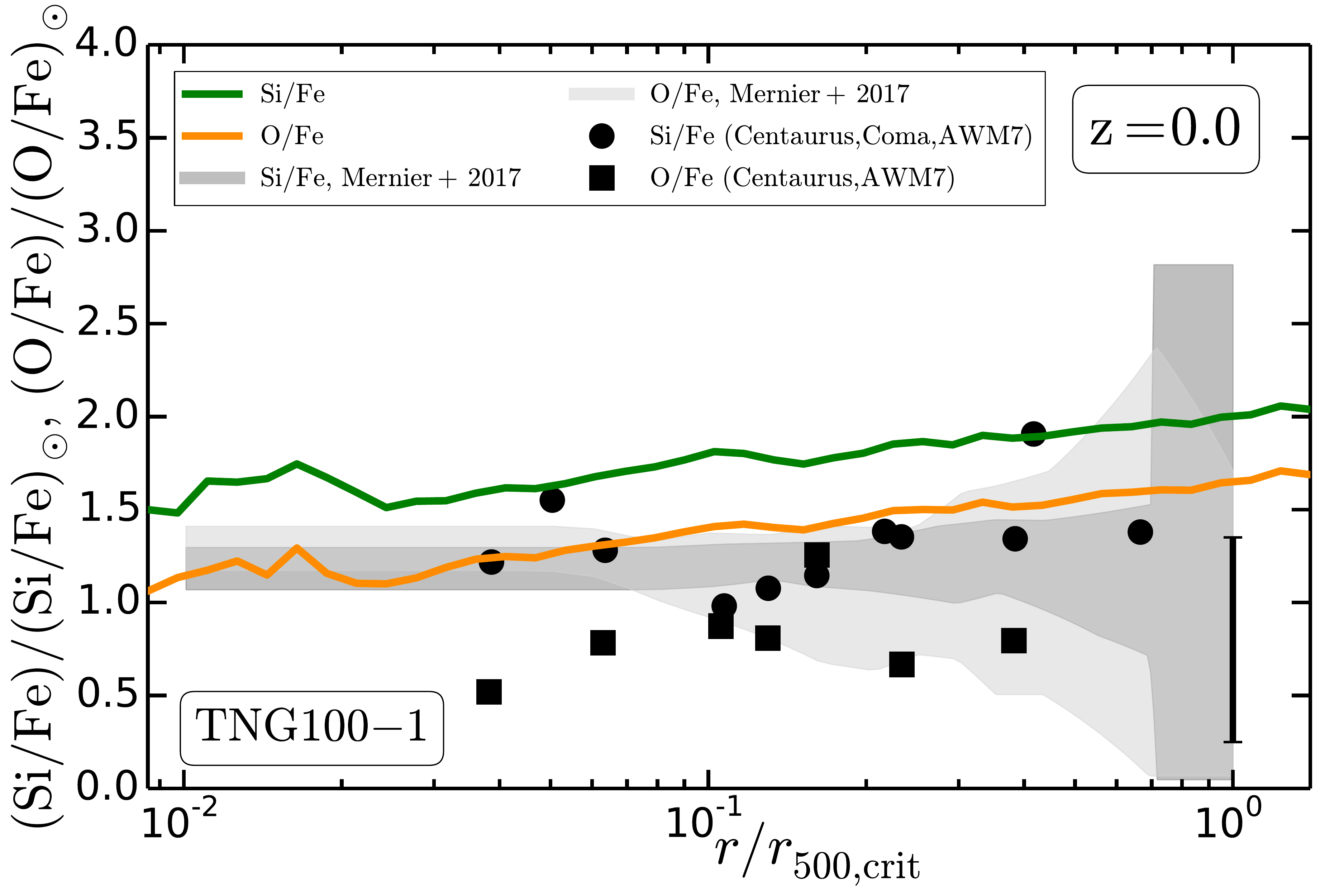}
\includegraphics[width=0.498\textwidth]{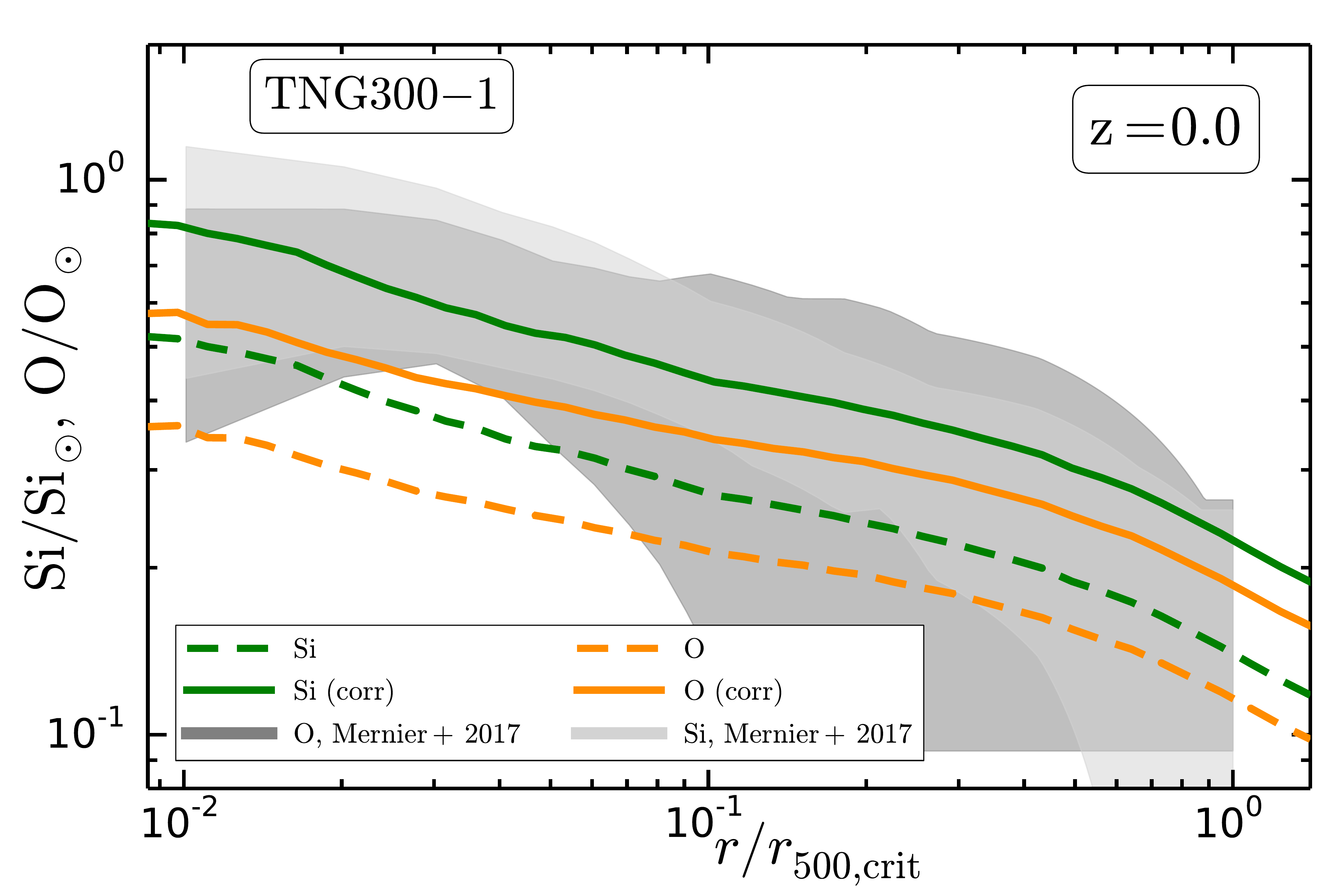}
\includegraphics[width=0.498\textwidth]{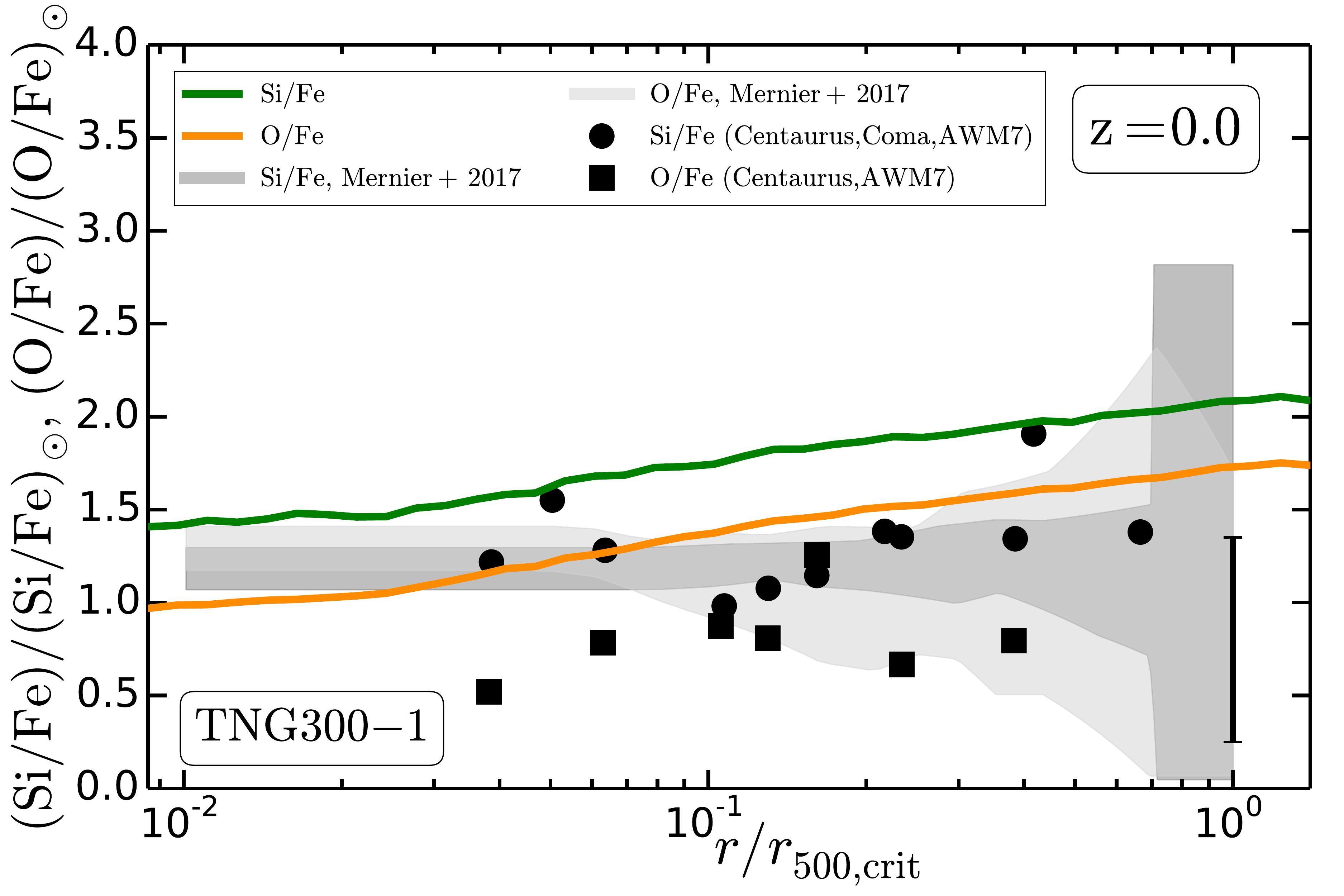}
\caption{Left two panels: Oxygen and silicon profiles at $z=0$ for TNG100-1
(top) and TNG300-1 (bottom). Lines show median value over the cluster samples
($M_{\rm 500, crit}>10^{13.75}\msun$) in both simulations. For TNG300-1 we
present also the resolution corrected profiles (corr). Observational data is
taken from \protect\cite{Mernier2017}. Right two panels: Abundance ratios of silicon and oxygen with respect to iron. The figure includes results for the
small TNG100-1 (top) sample and the significantly larger TNG300-1 (bottom)
cluster sample. Those ratios are resolution-independent and we therefore do not
apply any resolution corrections for TNG300-1. We include observational Suzaku
data points for AWM7 \protect\cite[][]{Sato2008}, Centaurus
\protect\cite[][]{Sakuma2011} and Coma \protect\citep[][]{Matsushita2013} (only
Si/Fe). The error bars on those measurements are large and vary with
cluster-centric distance. We indicate a typical abundance error in the lower
right of the panels. We also include more recent observational data from the stacking analysis of~\protect\cite{Mernier2017}. The simulation ratios are slightly too high compared to
the observational data. Furthermore, we also find that these ratios are declining
towards the cluster center. Both ratio decrease nearly by $0.5$ from $r_{\rm
500,crit}$ to $10^{-2}\times r_{\rm 500,crit}$.}
\vspace{-0.5cm}
\label{fig:element_profiles}
\end{figure*}

So far we have not yet divided our cluster samples into subsets according to
mass or cool core versus non-cool core for the analysis of the metallicity or
iron profiles. Observationally cluster metallicity profiles, average
metallicities and iron abundances do not seem to be strongly dependent on
cluster mass or temperature, which we will also demonstrate below. However,
observations do find that the metallicity profile depends on the thermodynamic
profile of clusters. Specifically, cool core clusters seem to have increased
metallicities at their centers compared to non-cool core clusters. These
differences in metallicity profiles for cool core versus non-cool core clusters
are only found for the innermost regions of clusters ($r < 0.15 \times r_{\rm
500,crit}$), where central enrichment increases the metallicity and element
abundances for cool core clusters.  Observationally cool core and non-cool core
clusters seem to have similar metallicity and iron profiles beyond that radius.
To explore this dichotomy in our simulations, we show in
Fig.~\ref{fig:cool_core} the median metallicity profiles for TNG100-1 divided
into cool core and non-cool core clusters.  We require a cool core cluster
to have a central entropy of less than $30\,{\rm keV}\,{\rm cm}^{-2}$~\citep[][]{Hudson2010}. The detailed
thermodynamical profiles of clusters require sufficient numerical resolution.
We therefore do not present a similar analysis for TNG300 due to its eight
times lower mass resolution. We compare the cool
core and non-cool core subsample profiles of TNG100-1 to observational data taken
from~\cite{Ettori2015}, where we select the low-redshift observations, $z<0.2$,
which we then compare to the $z=0.1$ predictions of the simulation. We note that their selection criterion
for cool core clusters is based on the pseudo-entropy ratio. The cool
core and non-cool core profiles of the simulation are in broad agreement with
the observational results and show a similar trend; i.e. cool core clusters
show rising metallicity profiles towards the center. Cool core clusters
approach $0.8\,{\rm Z}_\odot$ towards their centers whereas non-cool core
clusters have central metallicities below $0.5\,{\rm Z}_\odot$. We also find that the median metallicity profiles
of the cool core and non-cool core subsamples agree beyond $0.15 \times r_{\rm
500,crit}$ similar to the trends seen in observations. Dashed lines show the $1\sigma$ scatter for both the cool core
and non-cool core cluster samples.
We note that we have explored here only one cool core versus non-cool core
criterion based on the central electron density.  Other criteria like central
cooling time or entropy along with a more detailed census of cool core versus
non-cool core systems TNG300-1 is discussed in \cite{BarnesTNG}. We further note that the cluster sample in Fig.~\ref{fig:cool_core} is
rather small, but the simulation profiles indicate that the sample shows, at
least qualitatively, the observed dichotomy in the metallicity profiles.

Besides total metallicity and iron our galaxy formation model also traces other
elements as mentioned above. Here we
specifically focus on silicon and oxygen since they are, compared to iron, also
produced by SNcc on rather short timescales, whereas iron is mostly produced by
SNIa that operate on much longer time scales.  Differences in the iron versus
lighter element profiles therefore encodes information on the metal production
process and cluster formation history in general.  Observationally it appears
that the typical fraction of SNIa (SNcc) contributing to the enrichment lies
within $\sim 20 - 45\%$ ($55 - 80\%$), depending on the selected yield
models~\citep[][]{Mernier2017}. \cite{Bulbul2012b} finds a $30-37\%$ SNIa
fraction in the core of A3112. Observations further find that the silicon over
iron and oxygen over iron ratio distributions seem to be rather flat, which suggests a flat proportion and a similar metal contribution of SNcc and
SNIa~\citep[][]{Mernier2017}. 

Given the wealth of information encoded in the different element distributions,
we present in the left two panels of Fig.~\ref{fig:element_profiles} the median
profiles for silicon and oxygen. For TNG300 we show the raw profiles (dashed) and the
resolution corrected profiles (solid lines). The general shape of the silicon
and oxygen profiles coarsely follow those of iron and the total metallicity;
i.e. the abundance of these elements is increasing towards the center.  At
around ${\sim 0.1 \times r_{\rm 500,crit}}$ we find typical silicon abundances of
$\sim 0.4$ solar and oxygen abundances of $\sim 0.3$ solar. Those values drop
towards larger radii, $\sim r_{\rm 500, crit}$, down to $\sim 0.1 - 0.2 $ solar
for oxygen and silicon.  The general similarity of the oxygen and silicon
distribution compared to the overall metals and iron could already be seen in
Fig.~\ref{fig:element_maps}. We compare the detailed silicon and oxygen median
profiles to the recent stacking analysis of~\cite{Mernier2017} who studied $44$
nearby cool core galaxy clusters, groups, and ellipticals for their chemical
composition using deep XMM-Newton/EPIC observations to extract the average
abundance profiles of oxygen, magnesium, silicon, sulfur, argon, calcium, and
iron. We find that our results lie within the
scatter of measurements of the \cite{Mernier2017} observations. However, it
seems that the slope of the simulated oxygen and silicon abundances differs
slightly from those of the observational sample.

\begin{figure*}
\centering
\includegraphics[width=0.495\textwidth]{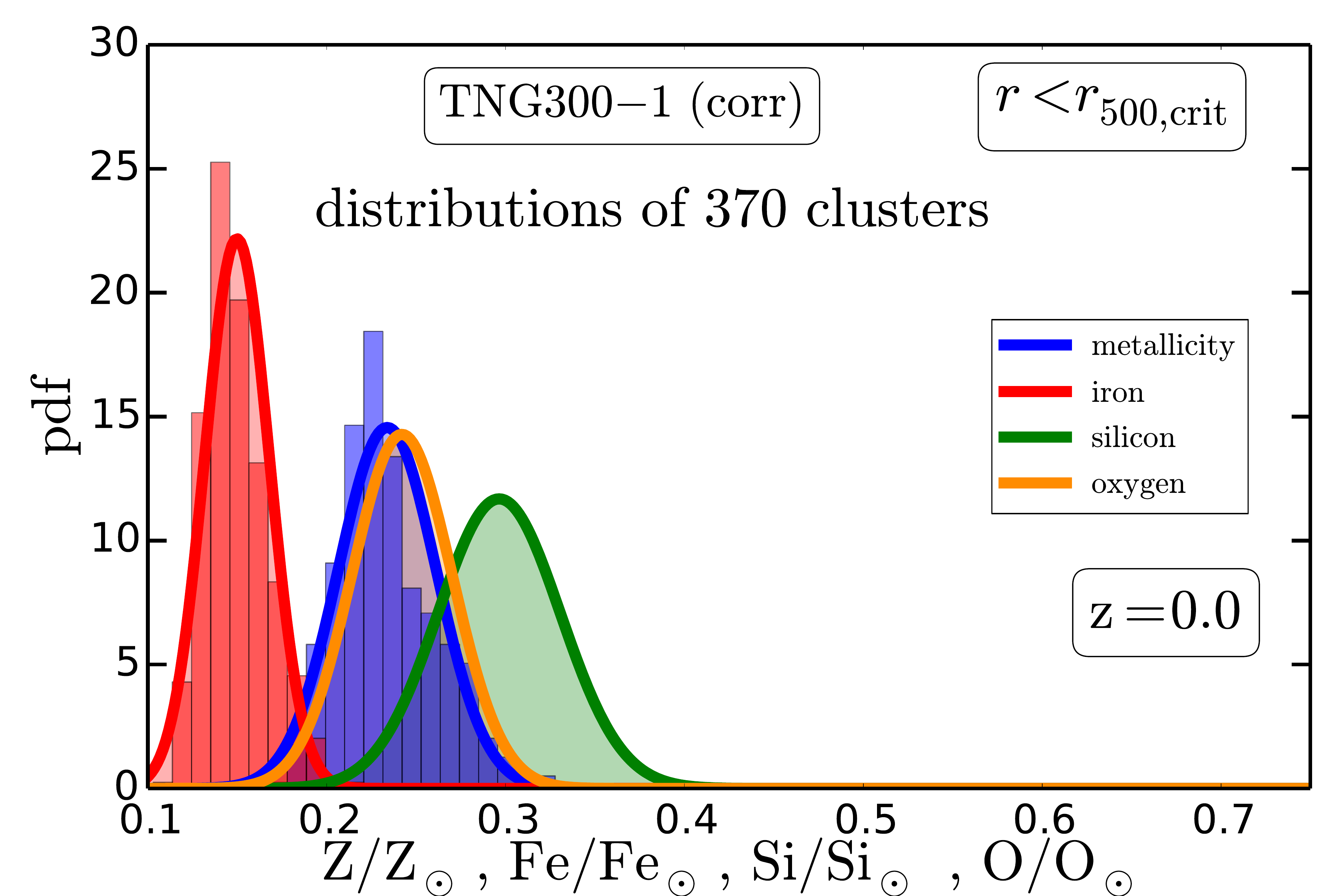}
\includegraphics[width=0.495\textwidth]{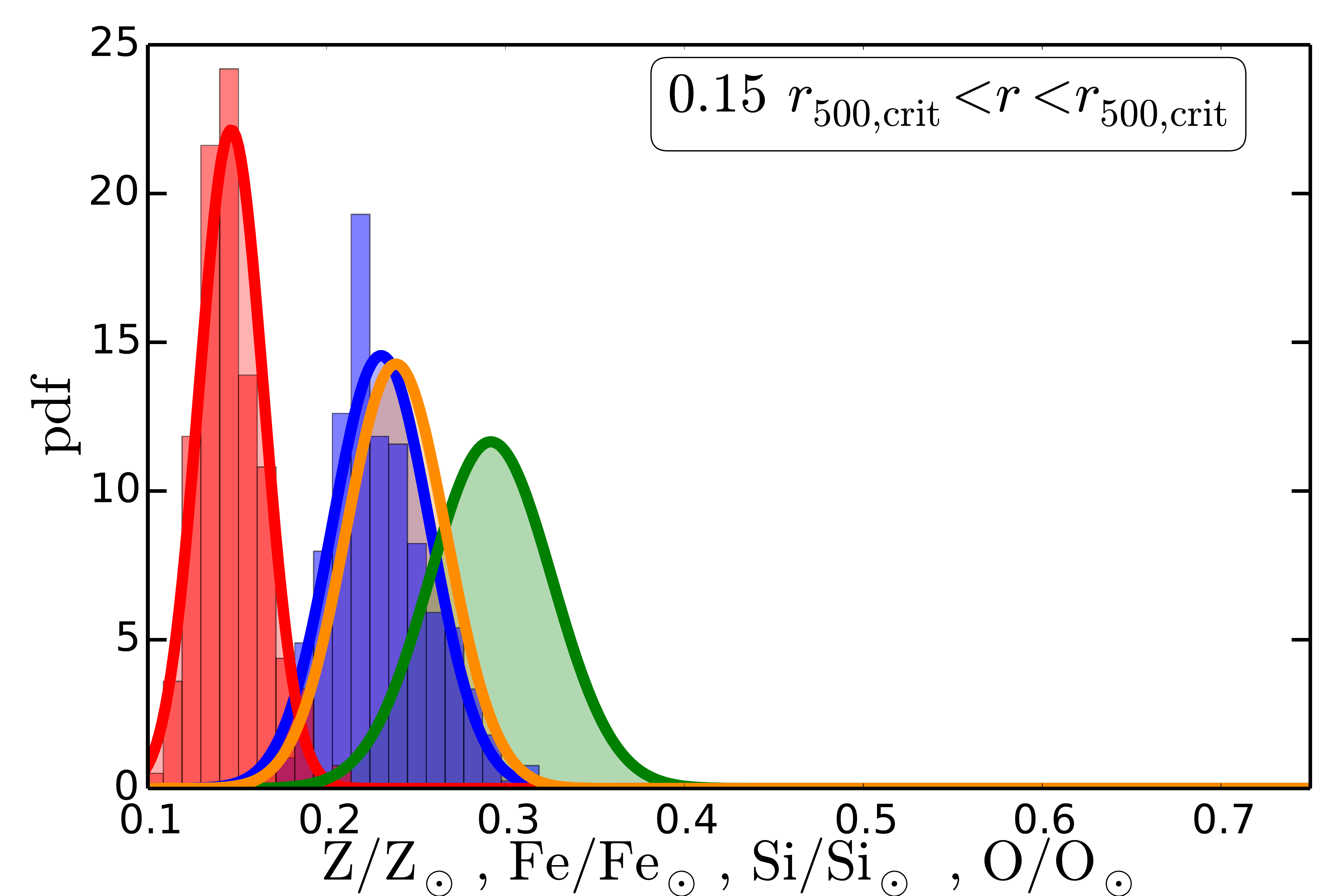}
\includegraphics[width=0.495\textwidth]{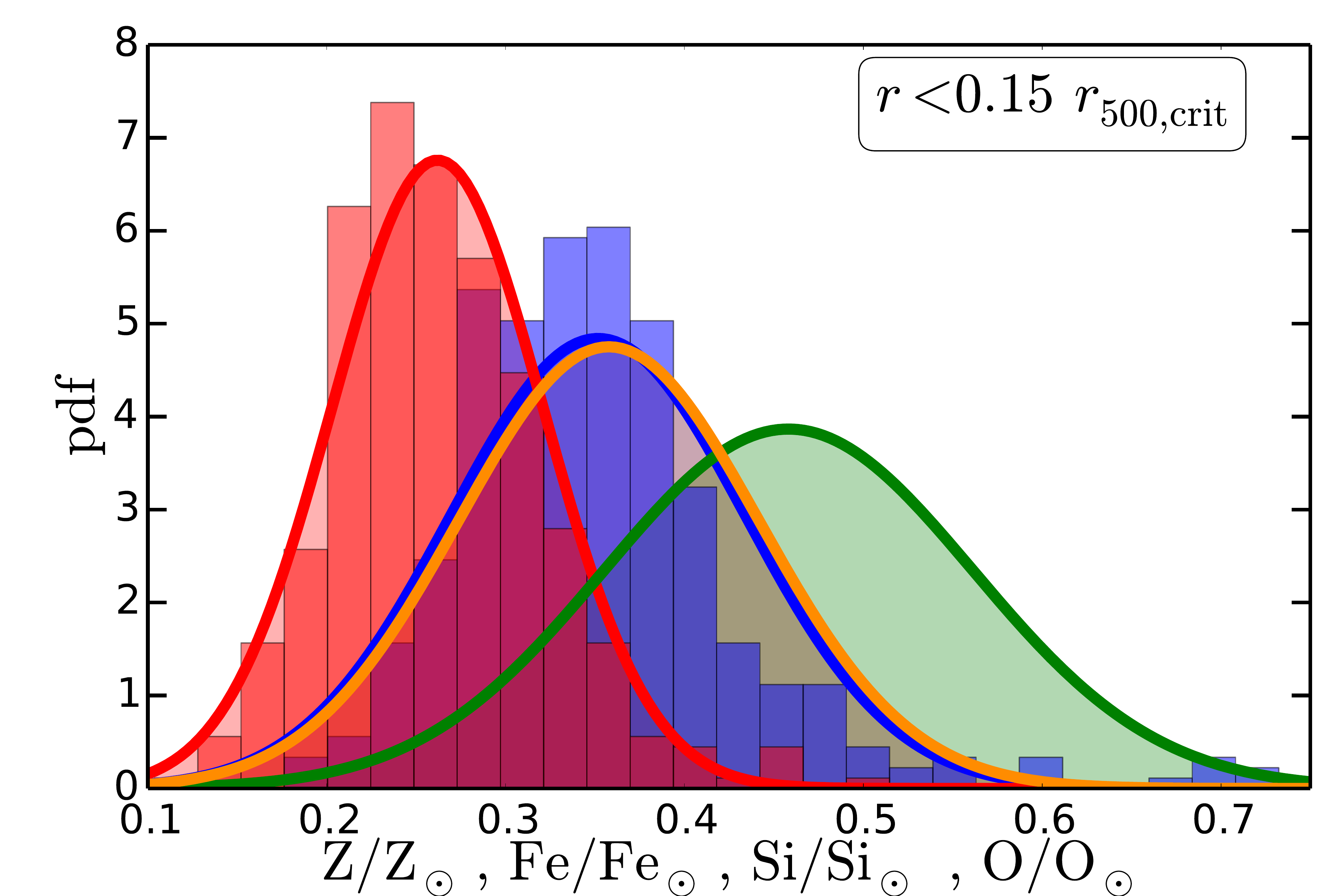}
\includegraphics[width=0.495\textwidth]{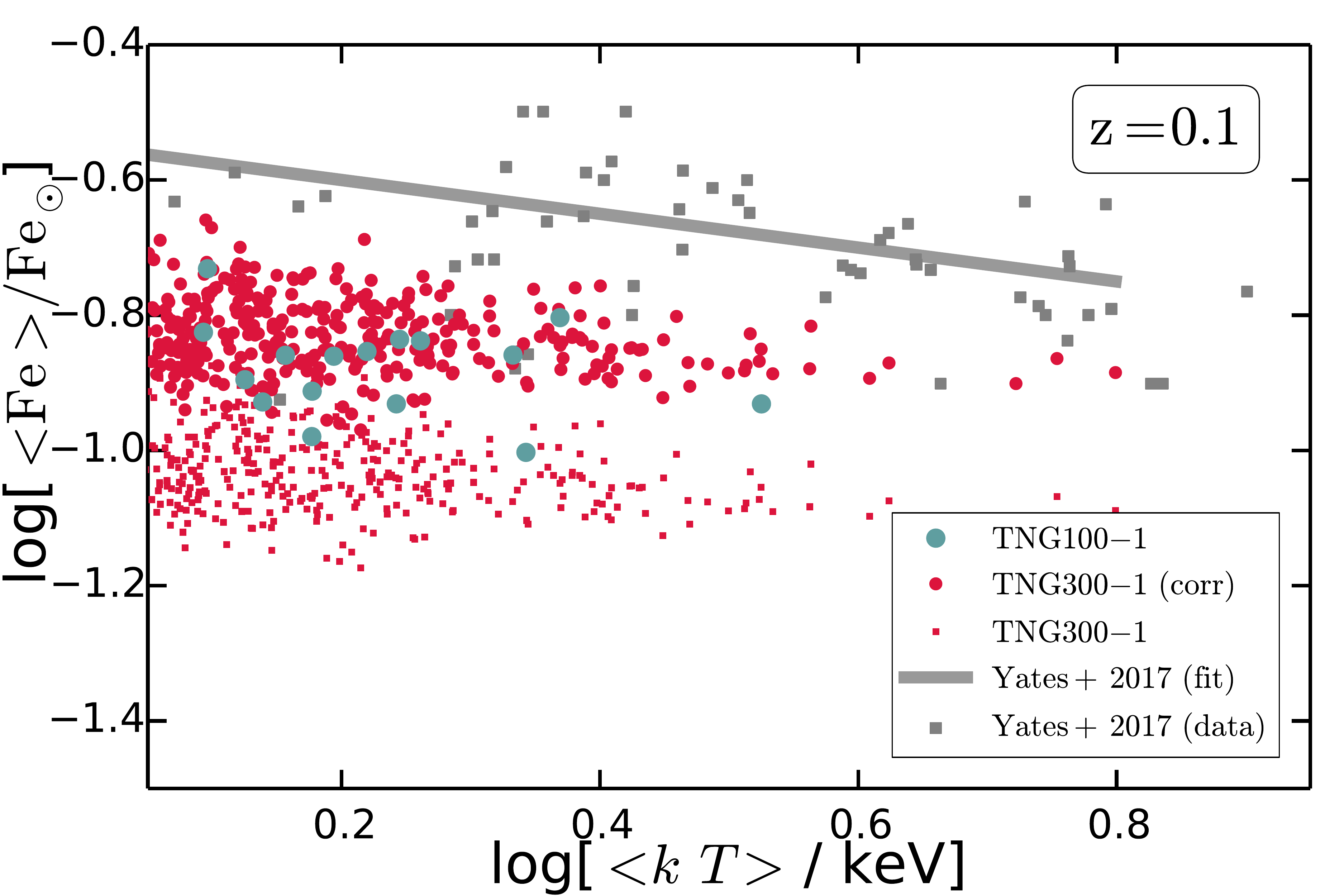}
\caption{Best-fit Gaussians for the mass-weighted average metallicities (Z),
iron (Fe), silicon (Si) and oxygen (O) abundances of our cluster sample measured within $r < r_{\rm 500,
crit}$ (upper left), $0.15 \times r_{\rm 500,crit} < r < r_{\rm 500,crit}$ (upper
right), and $r < 0.15 \times r_{\rm 500,crit}$ (lower left) for TNG300-1 with
resolution correction at redshift $z=0$.  The Gaussians have the following
$1\sigma$ dispersions in solar units for the different radial cuts ($r<r_{\rm 500,crit}$,
$0.15\times r_{\rm 500, crit}<r<r_{\rm 500, crit}$, $r<0.15\times r_{\rm 500,
crit}$): (Fe: $0.018$ Si: $0.034$ O: $0.028$ Z: $0.027$),  (Fe: $0.018$ Si:
$0.034$ O: $0.028$ Z: $0.027$), (Fe: $0.059$ Si: $0.103$ O: $0.084$ Z:
$0.082$). For metallicity and iron we also show the actual simulation data
demonstrating that Gaussians indeed describe those distributions well.  The
lower right panel presents the relation between average cluster temperature and
metallicity both for TNG100-1 and TNG300-1 (with and without resolution
correction)  where averages are calculated within $r_{\rm 500, crit}$.  In
agreement with observational data, there is no strong dependence of the iron
abundance on temperature over the full cluster sample; i.e. most clusters show actually
a similar amount of enrichment within $r_{\rm 500, crit}$. However, there seems
to be a slight anti-correlation which has also been pointed out
by~\protect\cite{Yates2017} with a similar slope as indicated (gray line). We also show the actual data taken from~\protect\cite{Yates2017} (gray square symbols).}
\vspace{-0.5cm}
\label{fig:element_pdf}
\end{figure*}
\begin{figure}
\centering
\hspace{-0.8cm}\includegraphics[width=0.52\textwidth]{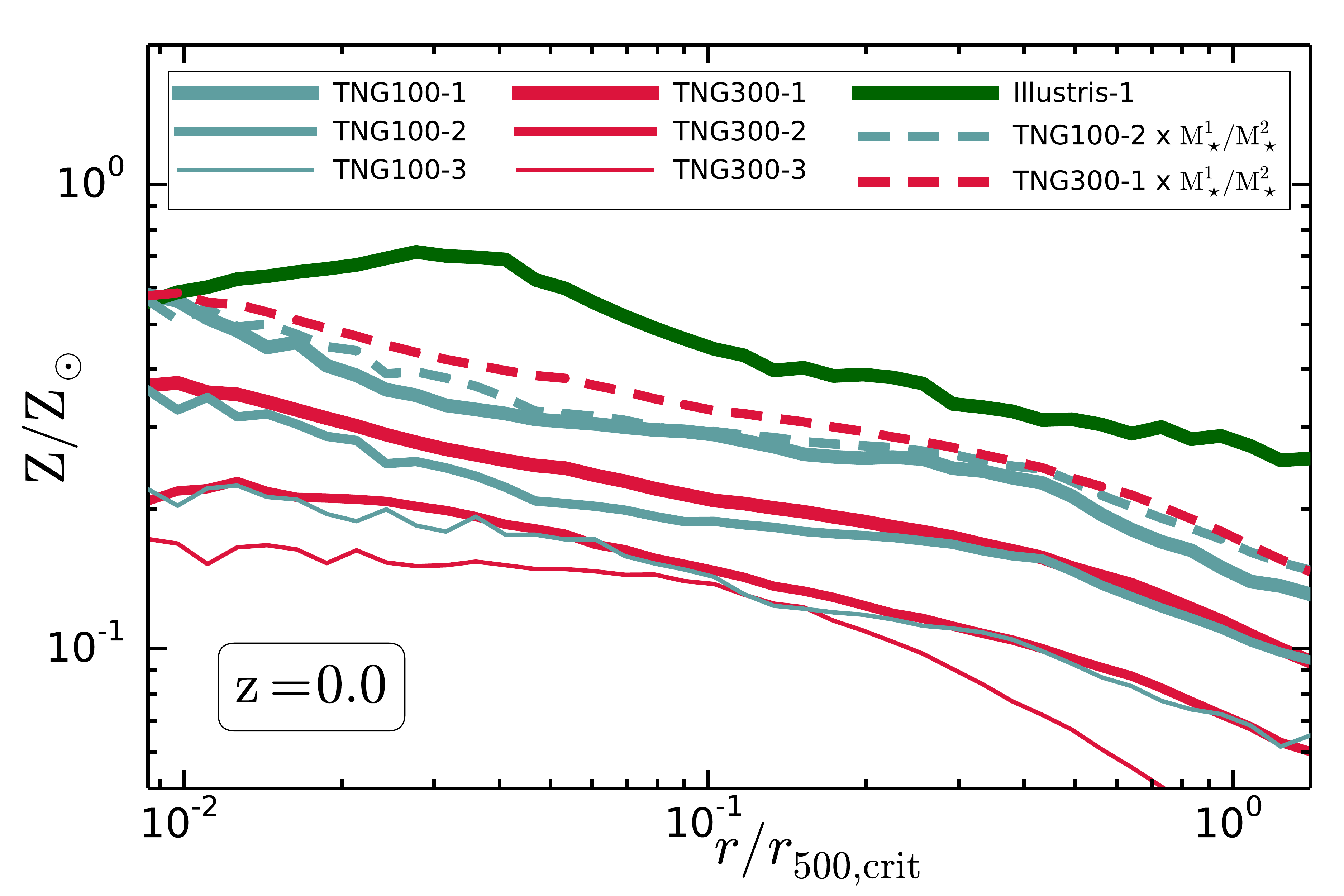}
\caption{Comparison of median cluster metallicity profiles extracted from Illustris and
IllustrisTNG, and TNG300/TNG100 at three different numerical resolutions.
Illustris and IllustrisTNG differ in the employed yields and details of the
feedback parametrisation, which impacts the ICM enrichment resulting in rather
different metallicity profiles.The Illustris galaxy formation model produces
central metallicity drops, which are absent in IllustrisTNG. We speculate that
this is related to the Illustris bubble radio-mode AGN feedback, which leads to
more efficient uplifting of central gas and metals compared to the kinetic
low-accretion rate AGN feedback of IllustrisTNG. The profiles of the different
simulation resolutions demonstrate that convergence for the ICM enrichment is
rather slow and it is typically difficult to achieve fully converged results.
However, a scaled up version of the TNG100-2 profile with a scaling factor
corresponding to the difference in stellar mass at level 1 and level 2
resolution ($M_\star^{1}/M_\star^{2}$) agrees with the TNG100-1 profile.  This
demonstrates that the reason for the slow convergence rate is ultimately due to
the slow convergence rate in stellar mass. Therefore, also the TNG300-1
metallicity profile overlaps with the median TNG100-2 profile, and a scaled up
version of TNG300-1 agrees approximately also with the profile of TNG100-1.
This resolution dependence motivates a simple resolution correction by scaling
up ICM abundances by $M_\star^{1}/M_\star^{2} \cong 1.6$ for TNG300-1 to the
resolution level of TNG100-1.}
\vspace{-0.5cm}
\label{fig:tng_vs_ill}
\end{figure}

To quantify this in more detail, we present in the two right panels of Fig.~\ref{fig:element_profiles} the
actual silicon over iron and oxygen over iron ratio profiles for TNG100 (top)
and TNG300 (bottom) in solar units. Since these are ratios, we do not have to
apply a resolution correction for TNG300. 
As in the other profile plots we
present here the median of the cluster samples. The ratios are
rather flat with a slight positive gradient, which shows that the overall
silicon and oxygen gradients are shallower than those of iron.  This
is consistent with Fig.~\ref{fig:metal_profiles}, where we found that the iron
profiles are steeper than those of the total metallicity.  We compare our
results to observational data from Suzaku for the massive galaxy clusters: AWM7
\cite[][]{Sato2008}, Centaurus \cite[][]{Sakuma2011} and Coma
\citep[][]{Matsushita2013} (only Si/Fe). We note that the observational error
bars are large, and the data is consistent with our simulation findings within
those errors.  We do not show the error bars for each data point, but instead
show the typical abundance ratio error in the lower right of each panel.
\cite{Mernier2017} also recently presented abundance ratio profiles, which we show for comparison.
In general, we find that oxygen is suppressed compared to silicon. Interestingly, the silicon
over iron and oxygen over iron ratios decrease slightly towards the
cluster center in our simulations. This is different from the ratio profiles
of~\cite{Mernier2017}, which are typically flatter in the inner part of the
cluster. In the simulation both ratios decrease by nearly $0.5$ from $r_{\rm
500,crit}$ to $10^{-2}\times r_{\rm 500,crit}$.

We note that some observational studies also find peaked oxygen
distributions~\citep[e.g.][]{Werner2006, Bulbul2012a}.  However, it is
difficult to reliably measure oxygen abundances. Oxygen lines are located at
$0.65$ and $0.56\,{\rm keV}$, and this band is heavily contaminated by soft
X-ray foreground emission coming from our own Galaxy. 

\begin{figure*}
\centering
\includegraphics[width=0.245\textwidth]{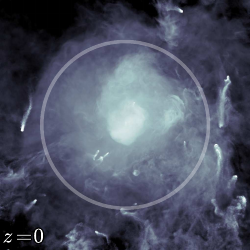}
\includegraphics[width=0.245\textwidth]{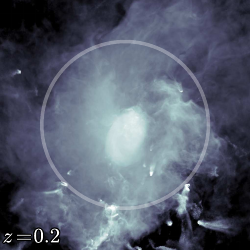}
\includegraphics[width=0.245\textwidth]{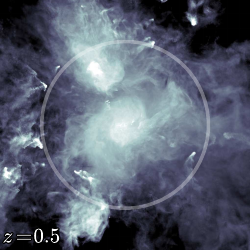}
\includegraphics[width=0.245\textwidth]{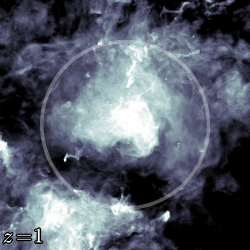}
\includegraphics[width=0.245\textwidth]{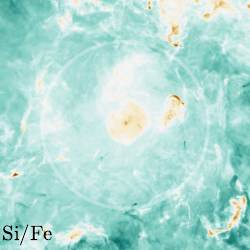}
\includegraphics[width=0.245\textwidth]{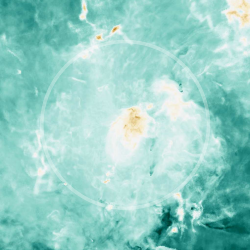}
\includegraphics[width=0.245\textwidth]{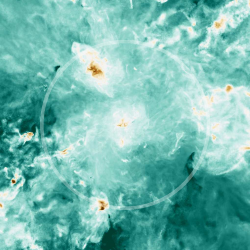}
\includegraphics[width=0.245\textwidth]{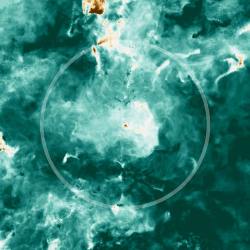}
\includegraphics[width=0.245\textwidth]{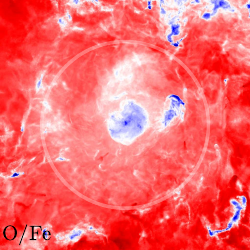}
\includegraphics[width=0.245\textwidth]{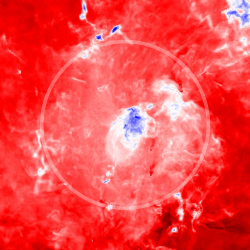}
\includegraphics[width=0.245\textwidth]{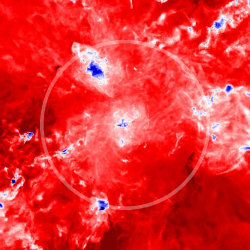}
\includegraphics[width=0.245\textwidth]{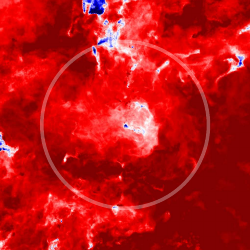}
\includegraphics[width=0.75\textwidth]{figures/maps/colorbar_fof_0_12_z0_Z_TNG300.pdf}\\
\includegraphics[width=0.25\textwidth]{figures/maps/colorbar_fof_0_z0_Si_over_Fe_TNG300_fof0.pdf}
\includegraphics[width=0.25\textwidth]{figures/maps/colorbar_fof_0_z0_O_over_Fe_TNG300_fof0.pdf}
\caption{Metallicity maps (top), Si/Fe maps (middle), O/Fe
maps (bottom) for the most massive clusters of TNG300 at $z=0$, $z=0.2$,
$z=0.5$, $z=1$. The extent and depth of each map is $3 \times r_{\rm 500,crit}$.
Colors are the same as in Figs.~\ref{fig:metal_maps} and
\ref{fig:element_maps}.  The average metallicity does not evolve much during
that time span, but the detailed metal distribution changes.}
\vspace{-0.5cm}
\label{fig:metal_maps_highz}
\end{figure*}

\subsection{Average metal content}

So far we have mainly studied radial abundance profiles. However, we can also
average the metallicity within clusters. We have already mentioned above that
this average metallicity of the ICM is observationally found to be quite
independent of the overall cluster properties like mass and temperature.
Specifically, the temperature-metallicity and X-ray-metallicity scaling
relations are found to be rather flat. Furthermore, the scatter in the average
metallicities is also quite small. We confirm these observational findings in
Fig.~\ref{fig:element_pdf} where we show best-fit Gaussians to the distribution
of mass-weighted average ICM metallicities, iron, silicon, and oxygen within
different radial cuts (first three panels) for TNG300 with the resolution
correction discussed above. For total metallicity and iron we also show the
actual data to demonstrate that the Gaussians provide a good fit.  The averages
for each cluster were computed in a mass-weighted way for $r<r_{\rm 500, crit}$
(upper left panel), $0.15\times r_{\rm 500, crit}<r<r_{\rm 500, crit}$ (upper
right panel), and  $r<0.15\times r_{\rm 500, crit}$ (lower left panel). In each
panel we can see a clear ordering of the abundance of elements with iron being
the least abundant and silicon being the most abundant in solar units. We also
find that the innermost abundances within $0.15\times r_{\rm 500,crit}$ are
higher than those in the outer parts of the ICM. The different distributions in
Fig.~\ref{fig:element_pdf} also demonstrate that the spread in the element
abundances over the full cluster sample of TNG300 is rather small showing the
uniformity of the enrichment process for the different clusters. The best fits
yield the following $1\sigma$ dispersions in solar units for the different radial cuts:\\[0.15cm]
$r<r_{\rm 500,crit}$:                                 Fe: 0.018 Si: 0.034 O: 0.028 Z: 0.027\\[0.1cm]
$0.15\,r_{\rm 500, crit}\!\!<\!r\!\!<r_{\rm 500, crit}$:   Fe: 0.018 Si: 0.034 O: 0.028 Z: 0.027\\[0.1cm]
$r<0.15\,r_{\rm 500, crit}$:                     Fe: 0.059 Si: 0.103 O: 0.084 Z: 0.082.\\[0.15cm]
In the lower right panel of Fig.~\ref{fig:element_pdf} we present the
temperature-iron abundance relation for TNG100-1 and TNG300-1, which
demonstrates that the average iron abundances of our simulation indeed do not
depend strongly on the average temperatures of the clusters in agreement with
observational findings. We calculate both the iron abundance and temperature as mass-weighted averages within $r_{\rm 500,crit}$. Actually there seems to be some slight
anti-correlation, which can also be seen in the recent compilation of
observational data presented in~\cite{Yates2017}. The gray line shows the
best-fit to this anti-correlation found by~\cite{Yates2017} using $79$
low-redshift ($z \cong 0.03$) systems resulting in a slope of $-0.26 \pm 0.03$, whereas points show the original data
points following the averaging procedure in~\cite{Barnes2017} taking mean
values if more than one estimate for the observed mass-weighted metallicity or
temperature was present in the~\cite{Yates2017} sample. The resolution
corrected values have been scaled up by $1.6$ to match the TNG100-1 resolution.
However, the simulation metallicities are slightly lower, $\sim 0.2\,{\rm dex}$, than the observational
values at our fiducial resolution. A similar discrepancy has also been found by
\cite{Barnes2017} when comparing the Cluster-EAGLE clusters with the
\cite{Yates2017} results.

\subsection{Model variations and convergence}

The results presented so far and the enrichment of the ICM depend on the actual
galaxy formation model that has been employed in the simulation.  The
IllustrisTNG galaxy formation model is an updated version of the previous
Illustris model as described above, and furthermore the TNG100 simulation retains all initial condition phases
of the original Illustris volume, but with an updated cosmology. We can
therefore directly compare the TNG100-1 results with those of Illustris-1 to
see how strongly the ICM enrichment varies between the two simulations. Such a
comparison provides important insights into the sensitivity of the ICM
enrichment on the galaxy formation model. In Fig.~\ref{fig:tng_vs_ill} we
therefore present a comparison between the median metallicity profiles derived
from Illustris (ILL-1) and IllustrisTNG (TNG100-1). We note that Illustris
contains only $10$ clusters fulfilling the mass cut criterion $M_{\rm 500,crit}
> 10^{13.75}\msun$ compared to the $20$ cluster within TNG100 of IllustrisTNG,
most importantly due to the different $\sigma_8$ normalisation.

The Illustris and IllustrisTNG simulations furthermore differ in various
aspects: (i) the yields have been changed between the two simulations, (ii) the
low accretion state AGN feedback works differently between the two simulations
with Illustris injecting thermal energy in inflating bubbles in the ICM, and
IllustrisTNG employing a kinetic feedback model injecting momentum at the
center of the cluster modelling a black hole driven wind, (iii) the
cosmology has been updated to the more recent Planck results for IllustrisTNG,
(iv) the details of the SN winds have been modified, (v) the metal advection
has been  improved in IllustrisTNG. All these changes affect the median
abundance profiles in the ICM, such that Illustris predicts nearly a factor
$\sim 1.5 - 2$ higher metallicity values compared to IllustrisTNG.  Aside from the
difference in normalisation we also see a difference in the shape of the
metallicity profile in the inner region of the ICM. The median profile of the
Illustris simulation is decreasing towards the center with a positive gradient.
IllustrisTNG, on the other hand, predicts a rising metallicity profile towards
the center with a negative metallicity gradient over the full radial range. We
speculate that this difference is related to the modelling of AGN radio mode
feedback. The radio mode feedback of Illustris drives gas out of the center of
cluster~\citep[][]{Genel2014} carrying along also metals. On the contrary, the
kinetic radio mode feedback of IllustrisTNG~\citep[][]{Weinberger2017a} does
not result in this gas removal and therefore also does not expel metals from
the central regions of the cluster. Besides this difference in the inner
part, we also find that the slope of the metallicity profile in the outer part
of the cluster, around and beyond $r_{\rm 500, crit}$, is shallower for the
Illustris model. In fact, the IllustrisTNG model shows a kink, which is absent
in Illustris. This could also be related to the difference in the AGN feedback
since the Illustris model distributes metals out to significantly larger radii
than the IllustrisTNG model.

Interestingly, there is also some observational evidence for decreasing
metallicities towards the center of clusters, and this has been interpreted as
uplifting of metals through AGN bubbles~\citep[e.g.,][]{Rafferty2013,
Kirkpatrick2015, Panagoulia2015, Mernier2017}. This seems also to be consistent
with the Illustris metallicity profile, where the profile indicates that
central metals have been moved to larger radii in the ICM.  \cite{Mernier2017}
found that $\sim 32\%$ of the clusters in their sample exhibit these kind of
metallicity drops towards the centers. 
For Illustris the median metallicity profiles
peak at $0.03\times r_{\rm 500, crit}$ reaching $\sim 0.7\,{\rm Z}_\odot$ and
drop to $\sim 0.6\,{\rm Z}_\odot$ at $10^{-2}\times r_{\rm 500, crit}$. This
is comparable to the central oxygen drops seen in~\cite{Mernier2017}, but larger
than the observed iron drop therein.
Metallicity drops could also be caused
by other physical effects, for example, dust-induced metal depletion
(Vogelsberger et al., in prep) or thermal conduction in the
ICM~\citep[][]{Kannan2017}. We note that the observed central metallicity drops
might also be a result of fitting simplified single-temperature models to a
multi-phase gas. Indeed, those drops often disappear if multi-temperature (2T
or 3T) plasma models are employed in the innermost
regions~\citep[e.g.,][]{Buote2003}.

We discussed above a few times the difficulty of numerical convergence of the ICM
element abundance profiles. To make this point clearer, we also present in
Fig.~\ref{fig:tng_vs_ill} the median metallicity profiles at all three
numerical resolution levels for TNG100 and TNG300. The mass resolution
differs by a factor of $\sim 64$ between level 3 and level 1. Higher numerical
resolution leads to more enrichment in the ICM reflected by an overall increase
of the normalisation of the metallicity profiles. Convergence for the ICM
enrichment is rather slow and is typically difficult to achieve~\citep[see also
the discussion in][]{Martizzi2016}.  Strictly speaking our results therefore
represent lower limits for the amount of metals in the ICM.  Interestingly, the
median metallicity profile of TNG300-1 overlaps with the median profile of
TNG100-2, and TNG300-2 overlaps with TNG100-3. It seems therefore that the
profiles at different resolutions can be scaled to each other with a constant
factor.

This is indeed possible because the normalisation offsets between different
resolution levels are mostly driven by differences in the amount of stellar
mass that is formed at the different resolution levels.  This can be seen by
looking at the profile of TNG100-2 scaled up with a constant factor
corresponding to the ratio between the stellar masses formed at the two
resolution levels ($M_\star^1/M_\star^2$). To calculate this ratio, we have
taken all clusters in the sample and summed up the stellar mass of each
associated friends-of-friends group. This then results in total stellar masses
$M_\star^{1,2}$ for TNG100-1,2, respectively.  The scaled version of TNG100-2
then overlaps with the metallicity profile of TNG100-1, which demonstrates that
the reason for the low convergence rate of the metallicity profiles is
ultimately due to the low convergence rate in stellar mass.  

\begin{figure*}
\centering
\includegraphics[width=0.495\textwidth]{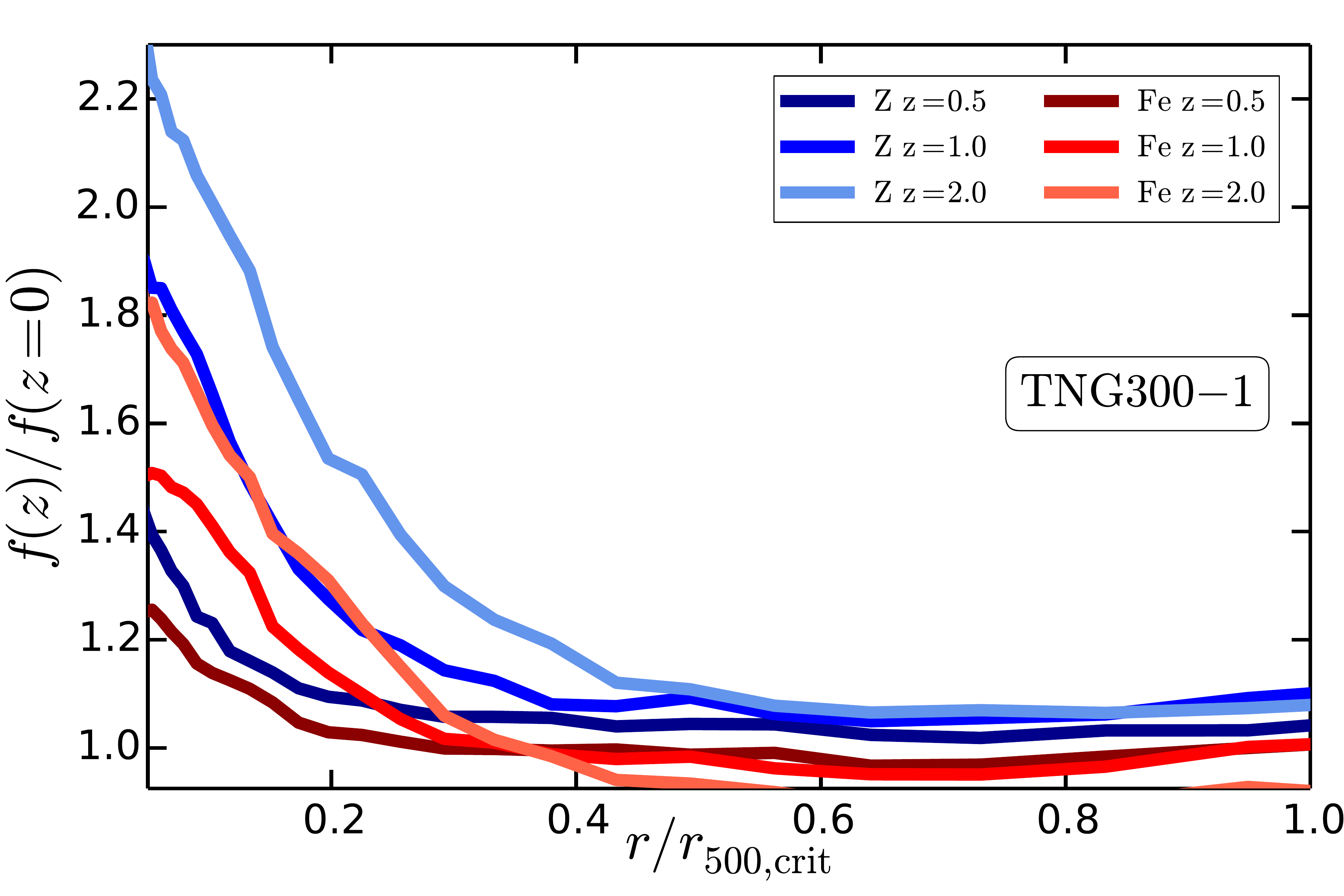}
\includegraphics[width=0.495\textwidth]{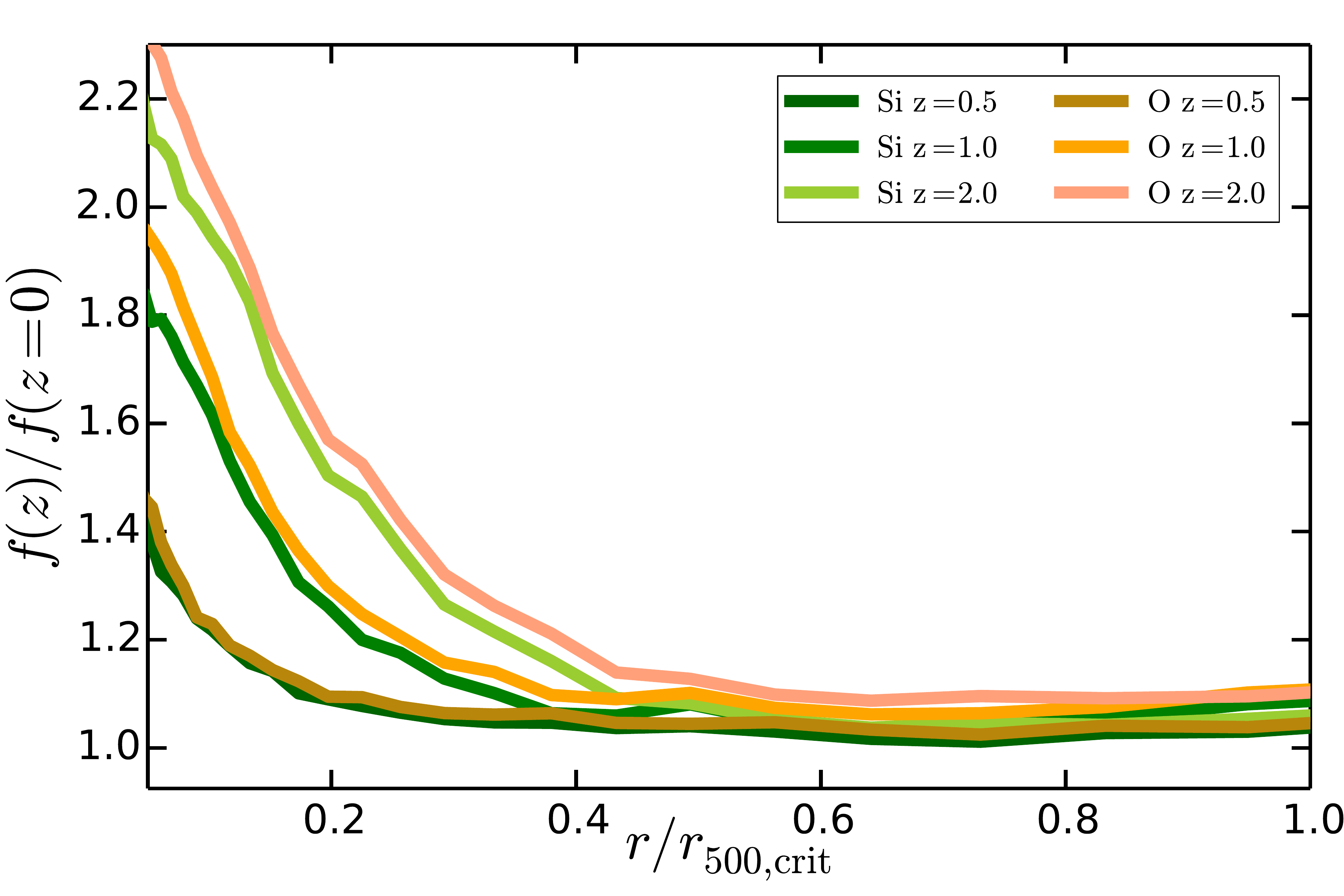}
\caption{Median metallicity, iron (left), silicon and oxygen (right) profiles
of TNG300 at $z=0.5, 1, 2$ compared to the $z=0$ profiles. The outer
parts of the ICM ($r \sim r_{\rm 500, crit}$) show a nearly constant level of
enrichment back to $z \sim 2$ changing by only about $\sim 10\%$. The slope of
the inner profiles ($r < 0.1 \times r_{\rm 500, crit}$) becomes shallower towards
$z=0$, and the central values decrease by nearly a factor of two from $z=2$ to
$z=0$. Therefore, the outer cluster abundance profiles beyond $0.4 \times r_{\rm 500,crit}$ are strikingly constant over time with nearly no evolution over the last $\sim 10\Gyr$.}
\vspace{-0.5cm}
\label{fig:metal_profiles_redshift}
\end{figure*}

We can use the scale factor $M_\star^1/M_\star^2 \cong 1.6$ between TNG100-1
and TNG100-2 to also scale up the profile of TNG300-1 to the mass resolution of
TNG100-1 as we have done above. This works reasonably well as
Fig.~\ref{fig:tng_vs_ill} demonstrates.  The difference in the profile shapes
between TNG100-1 and the scaled up version of TNG300-1 are most likely related
to the fact that the cluster sample of TNG300 is much larger and also includes
more massive clusters than the small sample of TNG100, which leads to a
slightly different shape of the median profile. Nevertheless, the fact that the scaled up
version of TNG300-1 agrees so well with TNG100-1 demonstrates that stellar mass
is the main driver for non-convergence, and the uniformity of metal enrichment
among different clusters is the reason for the approximate similarity of the
profile shapes that allows a resolution scaling solely based on the stellar
mass ratio.  One might ask whether the normalisation differences between
Illustris and IllustrisTNG are also mostly driven by different amounts of
stellar mass. We have checked that this is not the case since the stellar
mass within the cluster sample is only $\sim 10\%$ higher in Illustris
compared to IllustrisTNG, which is not sufficient to explain the offset.
Furthermore, the profile shapes are obviously different such that a simple
scaling cannot map one profile to the other.

The results presented in this section demonstrate that the ICM enrichment of IllustrisTNG
clusters is in good agreement with observational data. However, we note that
the silicon over iron and oxygen over iron abundance ratios are in slight
tension with observations since they follow a shallow positive gradient which
is disfavored by current observations pointing towards flatter profiles.

\section{Past ICM Composition and its evolution}\label{sec:past_icm}

In the previous section we have focused on the present-day metal, iron, silicon
and oxygen abundances within the ICM. One key result is the uniformity of these
profiles across the large cluster samples that we studied. Additionally,
observations also find only little evolution of the metal content of galaxy
clusters as a function of time~\citep[e.g.,][]{Ettori2015, McDonald2016,
Mantz2017}.  Therefore, both the variability from cluster to cluster and also
the variability as a function of time seem to be rather small in terms of the
ICM metal content. Observationally obtaining reliable radial metallicity
profiles for nearby clusters is challenging. Performing this exercise at higher
redshifts is even more uncertain. However, average iron abundances and
metallicities can be extracted for reasonably large samples of clusters
spanning significant lookback times. \cite{McDonald2016} analysed cluster
observations out to $z \sim 1.5$ finding nearly no evolution of the averaged
cluster metallicities within $r_{\rm 500, crit}$ as a function of redshift, in
contrast to results of~\cite{Ettori2015} who found a mild redshift dependence.
However, in both of these studies the evolution seems to be mostly driven by
changes in the actual core region with little evolution in the outer
cluster parts beyond $0.15 \times r_{\rm 500, crit}$.  Specifically, both
studies do not find a significant redshift dependence once the core region is
excluded, in which case they also do not find any significant trend with cool
core versus non-cool core thermodynamic profiles.  More
recently~\cite{Mantz2017} analysed the redshift evolution of the iron content of
$254$ massive clusters based on Chandra and Suzaku observations also finding essentially no evolution especially in the
outer parts of the ICM.  In summary it seems that observationally the outer ICM
metallicity profiles do not strongly evolve with redshift, while the inner parts show some evolution,
although not all observational results are fully consistent with each other. In this section we
will study the enrichment evolution of our cluster sample in more detail. 

To begin we first show qualitatively the evolution of the metallicity maps of
the most massive cluster of TNG300 in Fig.~\ref{fig:metal_maps_highz}. At each
redshift the extent of the projection is $3 \times r_{\rm 500,crit}$, and as in
the maps in the previous section the circles denote $r_{\rm 500,crit}$.  The
mass-weighted average metallicity (resolution corrected) in solar units within
this radius changes from $0.127$ ($0.203$) at $z=0$, to $0.124$ ($0.198$) at
$z=0.5$, and to $0.138$ ($0.221$) at $z=1$. Therefore, the overall metal budget
is not strongly evolving with redshift for more than $\sim 7\Gyr$.  However, local changes in the actual
distributions of the metals in the ICM are visible in the maps. We note that
this is in agreement with the observational findings of~\cite{McDonald2016};
i.e. the average metal content does not evolve below $z \sim 1.5$. At some
redshifts, for example at $z=0.5$, one can also see how mergers bring in metal
rich gas. The second and third row of Fig.~\ref{fig:metal_maps_highz} show the
corresponding time evolution of the silicon over iron and oxygen over iron
abundance ratios. 

\subsection{Redshift evolution of metallicities}

To quantify the redshift evolution of the ICM enrichment in more detail, we
present in Fig.~\ref{fig:metal_profiles_redshift} the median metallicity, iron,
silicon and oxygen profiles of the cluster samples of TNG300 at different
redshifts ($z=0.5$, $z=1$, $z=2$) compared to the present-day profiles by showing their ratios. Here,
and in the following, we trace back the clusters in our sample and look at
their progenitors at different redshifts.  We find that the outer parts of the
ICM do not evolve much since $z=2$, and the profiles agree well with those at
$z=0$.  Therefore the outer parts ($\sim r_{\rm 500, crit}$) of the ICM are not
only very universal among different clusters, reflected by the small scatter at
larger radii found in the previous section, but they are also invariant over a
long period of time. Hence, enrichment does not seem to occur in those outer
parts and any evolution is restricted to the more central parts of the cluster.
In fact, the outer cluster abundance profiles beyond $0.4 \times r_{\rm
500,crit}$ are strikingly constant over time with nearly no evolution over the
last $\sim 10\Gyr$.  This finding is in agreement with other theoretical
studies~\citep[e.g., ][]{Biffi2017} and is also consistent with observational
results as described above. The median profiles in
Fig.~\ref{fig:metal_profiles_redshift} also demonstrate that there is some
evolution of the profiles occurring at the centers of the clusters, where they
tend to become flatter towards later times.  The central amplitudes decrease by
nearly a factor of $2$ at the cluster centers from $z=2$ to $z=0$. We note that
\cite{Martizzi2016} found a similar effect in their simulations, but
this significant evolution seems to be absent in observations~\citep[e.g.,][]{Mantz2017}. It is
unclear whether this points to a systematic uncertainty in the
observations, or whether the central enrichment in simulations, and related to
this potentially the AGN feedback, are inconsistent with the observed central metallicity evolution at cluster
centers. 

\begin{figure}
\centering
\hspace{-0.25cm}\includegraphics[width=0.49\textwidth]{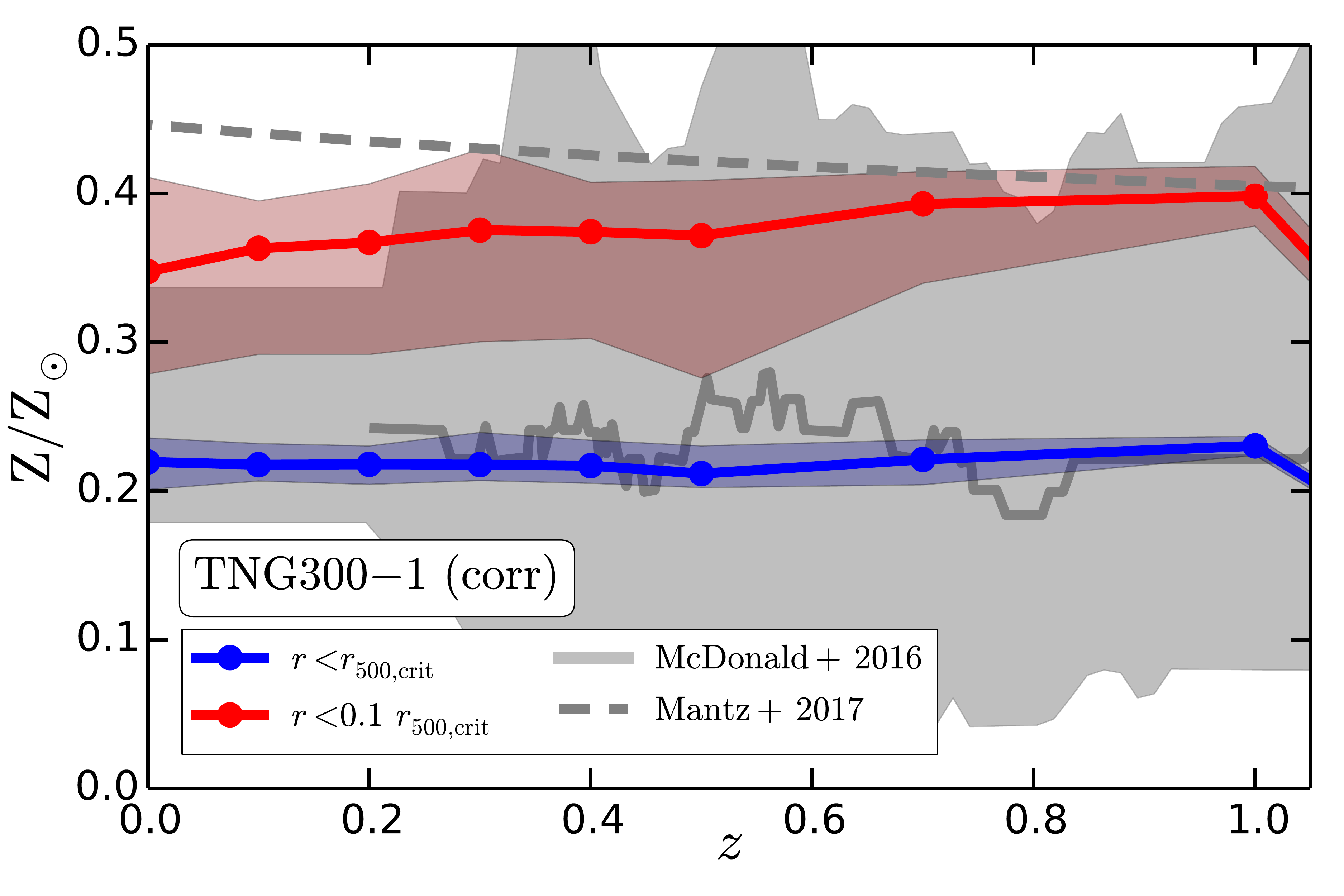}
\caption{Redshift evolution of the mass-weighted average metallicities for
clusters of TNG300 with resolution correction. We
present the redshift evolution within two different radial cuts: inner ($r<0.1 \times r_{\rm 500, crit}$), and total ($r<r_{\rm 500, crit}$). For the total cut
the median over the cluster sample does not evolve significantly with redshift.
For the inner regions we find that the metallicity slightly decreases towards lower redshifts
consistent with the findings in Fig.~\ref{fig:metal_profiles_redshift}.  The
constancy of the total metallicity is in agreement with recent observational
findings as indicated by the gray median line and shaded region taken
from~\protect\cite{McDonald2016} also measured within $r_{\rm 500, crit}$. We
also compare the evolution of the central average iron abundance to the best-fit
model of~\protect\cite{Mantz2017}. The
bands for observations and simulation data show the $1\sigma$ scatter. This is much larger for the
observational data since the cluster to cluster variation is dominated by
measurement uncertainties. Based on our simulations we find that the true cluster to cluster
variation is significantly smaller, $\sim 0.03\,{\rm Fe}_\odot$, for all
redshifts for the total cut ($r<r_{\rm 500, crit}$). 
}
\vspace{-0.5cm}
\label{fig:median_Z_redshift}
\end{figure}

Besides studying median abundance profiles we can also average the
metallicity over the full cluster within $r_{\rm 500,crit}$ as we have done in the previous section. We can then
calculate medians and percentiles of all these averages to inspect the redshift
evolution of the median metallicity of our cluster sample.  This procedure
mimics the observational analysis of, for example, \cite{McDonald2016} and \cite{Mantz2017}. We
present the resulting evolution in Fig.~\ref{fig:median_Z_redshift}. At each
redshift we neglect cluster progenitors below $M_{\rm 500,crit } < 2 \times
10^{14}\msun$ when calculating the medians to be consistent with the
observational samples of~\cite{McDonald2016} and \cite{Mantz2017}. Otherwise a large fraction of the
progenitors at higher redshift would drop out of the range of observed clusters analysed
in~\cite{McDonald2016} and \cite{Mantz2017} and we would get a median evolution that is biased
towards lower masses.  The mass cut of $M_{\rm 500, crit} \ = 2 \times 10^{14}\msun$ is approximately
the lower limit of clusters entering the median of \cite{McDonald2016} and \cite{Mantz2017} until a
redshift of $1.7$. We note however that we miss, especially at higher
redshifts, a fraction of massive clusters in the samples of~\cite{McDonald2016} and~\cite{Mantz2017}
since the simulation volume of TNG300 is too small to host those. For example,
the sample of \cite{McDonald2016} contains at $z \sim 1$ a few clusters with
masses larger than $M_{\rm 500, crit}=10^{15}\msun$, which are not present in
TNG300 due to the limited simulation volume.

In Fig.~\ref{fig:median_Z_redshift} we present the median evolution of the metallicity
averages within two different radial regions within the cluster: inner ($r<0.1
\times r_{\rm 500, crit}$), and total ($r<r_{\rm 500, crit}$). Our findings for
the redshift evolution of the cluster sample median is in qualitative agreement
with the results of \cite{McDonald2016} who found for $153$ galaxy clusters
between $0<z<1.5$ only a very weak redshift dependence for the average
metallicities within $r_{\rm 500,crit}$.  Specifically, \cite{McDonald2016} fit
their median metallicity redshift dependence within $r_{\rm 500, crit}$ with a
linear model $Z(z^\prime)=a + b z^\prime$ where $z^\prime = z
- 0.6$ is chosen to roughly minimise the covariance between $a$ and $b$ for the
  subsample of clusters observed with Chandra ($z\sim 0.6$). For the global
metallicity within $r_{\rm 500,crit}$ they find $a=0.23 \pm 0.01\,Z_\odot$ and
$b=-0.06 \pm 0.04\,Z_\odot$. Within the error bars this is consistent with
nearly no evolution. We note that this overall redshift dependence is weaker
than the one reported in~\cite{Ettori2015}.  However, the analysis
of~\cite{McDonald2016} presented the first study of the ICM metallicity evolution in a
sample of galaxy clusters selected via the Sunyaev-Zel'dovich effect resulting in a selection
that represents a mass-limited, redshift-independent sample that is relatively
unbiased with respect to the presence or lack of a cool core. This sample is
therefore significantly less biased than previously employed cluster samples to
explore the redshift evolution of the average metallicity of clusters.
Correspondingly, the  X-ray follow-up of galaxy clusters at $0 < z < 1.2$
provides therefore excellent constraints on the global metallicity evolution and
the radius-dependent evolution over this redshift range. \cite{Mantz2017} presented recently a similar even more complete study. Overall we find that
the median mass-weighted averages agree reasonably well with the observational
median of~\cite{McDonald2016} ($r < r_{\rm 500, crit}$) and the best-fit evolutionary model of~\cite{Mantz2017} ($r < 0.1 \times r_{\rm 500, crit}$). There is a slight redshift evolution in the simulation data, which
might be caused by the limited number of progenitors towards higher redshifts.
We also note that the redshift evolution in the central region ($r<0.1 \times
r_{\rm 500, crit})$ is stronger, showing an increase of about $\sim 20\%$ from $z=0$ to $z \sim 1$, in agreement with the evolution of the radial
profiles presented in Fig.~\ref{fig:metal_profiles_redshift}. However, this trend is not seen in
the observational data although the average enrichment level agrees with observations. 
Fig.~\ref{fig:median_Z_redshift} also includes bands showing the $1\sigma$
scatter over the simulated cluster samples. This scatter is much larger for the observational data since the
cluster to cluster variation is dominated by measurement uncertainties.
Based on our simulations we find that the true physical scatter for $r<r_{\rm 500, crit}$, $\sim 0.03\,{\rm Fe}_\odot$ at all redshifts, is significantly
smaller than the observational variation.

\begin{figure}
\centering
\hspace{-0.25cm}\includegraphics[width=0.49\textwidth]{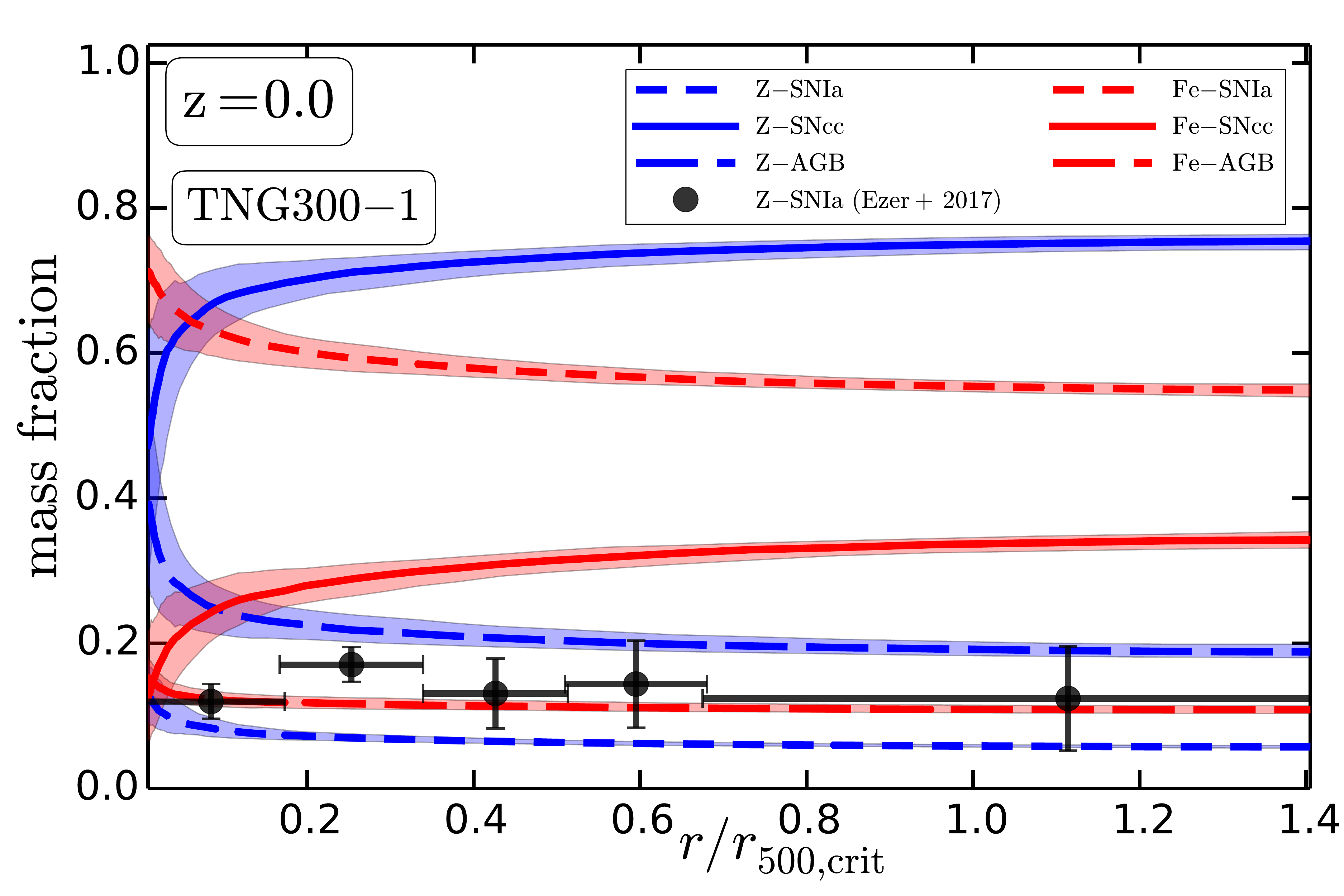}\\
\hspace{-0.25cm}\includegraphics[width=0.49\textwidth]{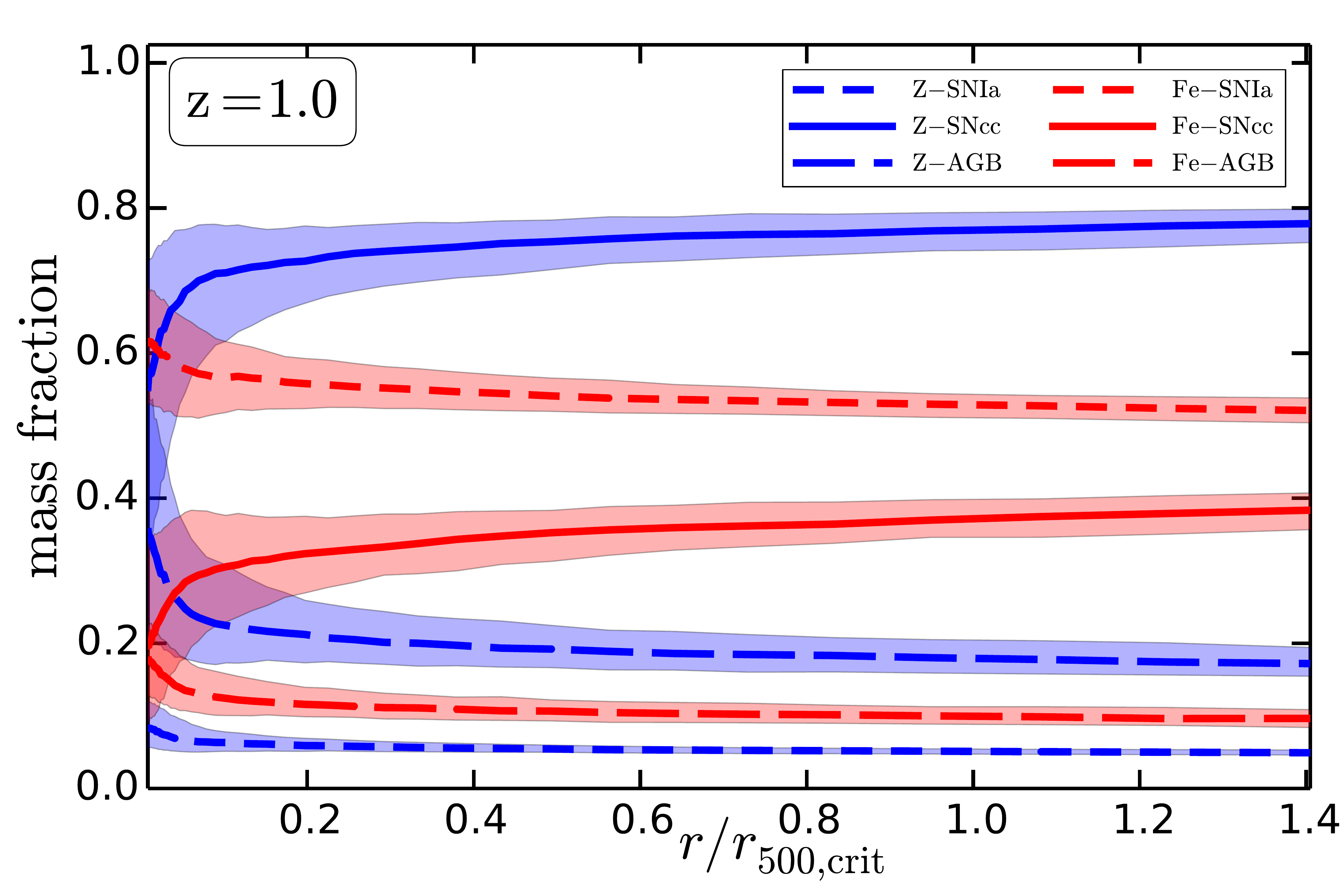}\\
\hspace{-0.25cm}\includegraphics[width=0.49\textwidth]{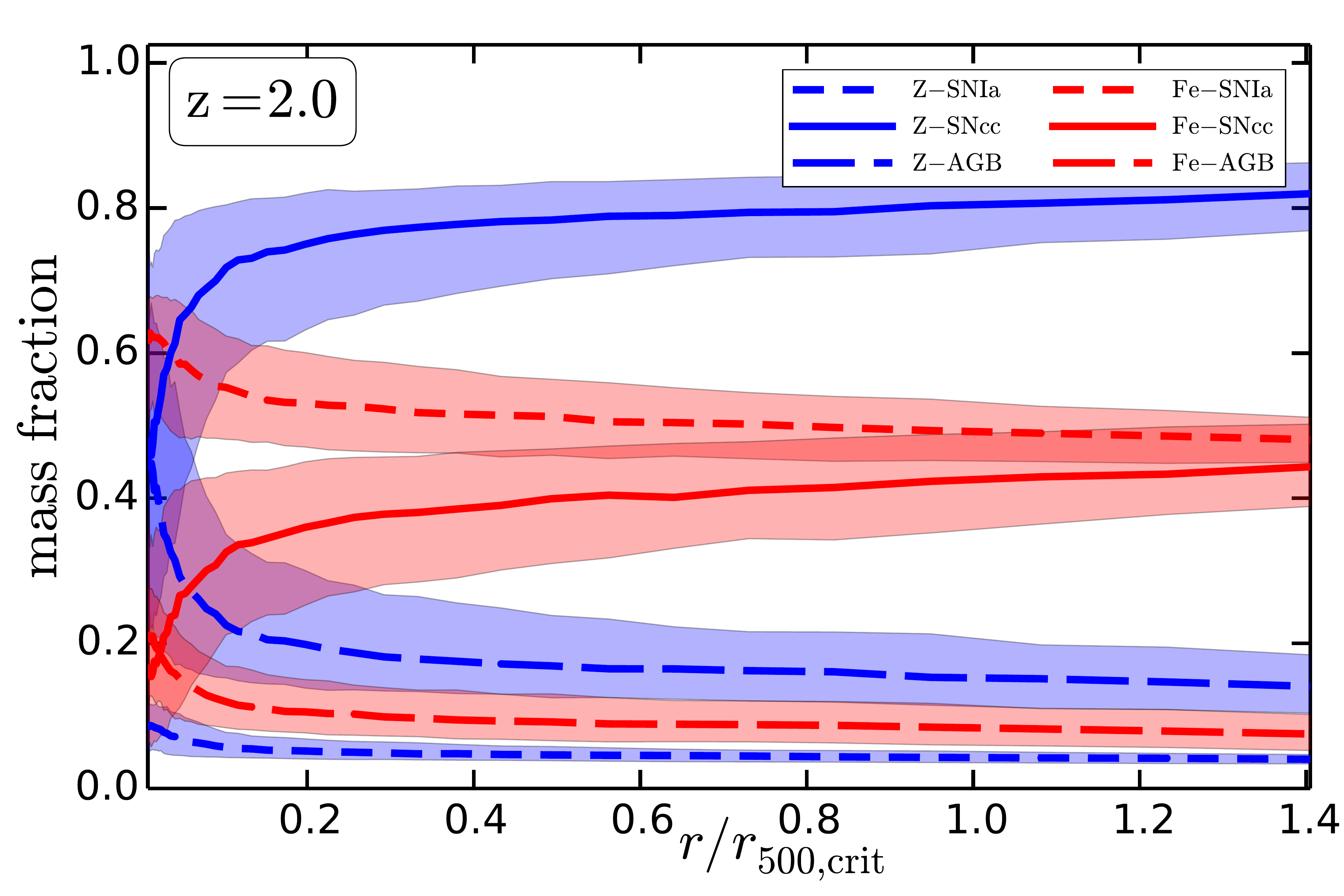}
\caption{Median radial profiles of the production sources of iron and metals split in
three different production channels: mass fraction resulting from AGB, SNIa,
SNcc for TNG300 at different times ($z=0, 1, 2$). Shaded regions show the
$1\sigma$ scatter over the cluster sample. The iron budget is at all radii
dominated by production through SNIa, whereas SNcc dominate the overall metal
budget in the ICM. The observationally inferred SNIa mass fraction is based on
the SNIa versus SNcc rates presented in~\protect\cite{Ezer2017}. Those are
nearly consistent with the simulation results within the observational error
bars. We find at all redshifts that the metal contribution from SNcc is slowly
decreasing towards the cluster center, whereas the SNIa is very slowly
increasing. As a consequence the silicon over iron and oxygen over iron ratios
are also decreasing towards the center as demonstrated in
Fig.~\ref{fig:element_profiles}.}
\vspace{-0.5cm}
\label{fig:Z_tagging}
\end{figure}

\begin{figure}
\centering
\hspace{-0.25cm}\includegraphics[width=0.49\textwidth]{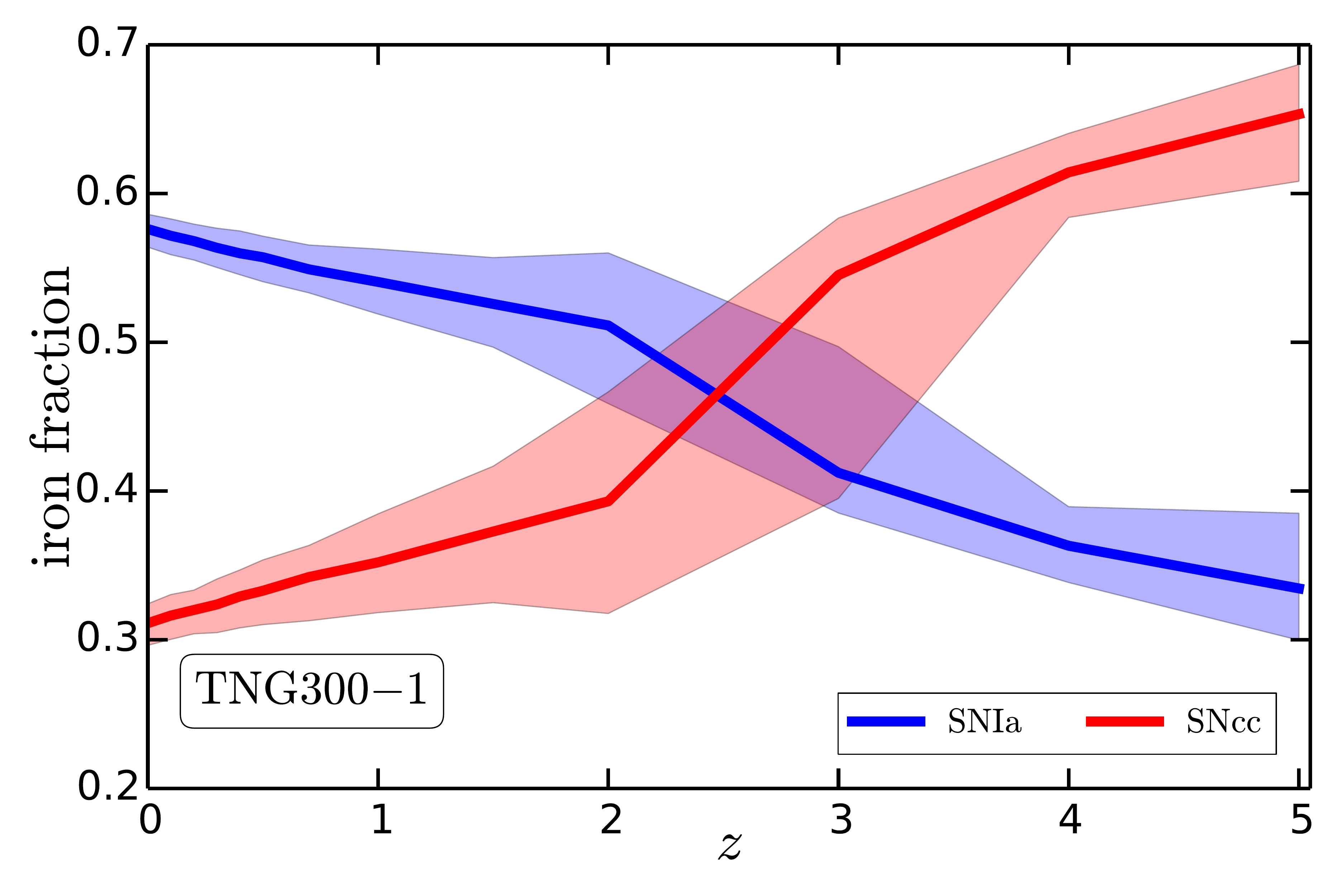}
\caption{The fractional mass contribution of SNIa and SNcc to the total iron budget
within $r_{\rm 500,crit}$ at different redshifts. At early times SNcc
contribute most of the iron until $z~\sim 2.5$ since SNcc enrichment occurs on
shorter timescales than SNIa enrichment. At $z\sim 2.5$ the SNIa contribution
begins to dominate the overall iron budget. The transition is rather quick such
that below $z \sim 2$ the contribution of SNIa versus SNcc is quite constant in
agreement with the profiles presented in Fig.~\ref{fig:Z_tagging} below $z =
2$.}
\vspace{-0.5cm}
\label{fig:median_tag_redshift}
\end{figure}

\subsection{Origin of metals}

The metal and element budget presented above originates from three different
stellar population sources: AGB stars, SNIa, and SNcc. We can trace back the
origin of metals and iron at different radii within the ICM using the metal
tagging technique of {\sc Arepo}. The resulting fractional mass
profiles with the different metal and iron sources are presented in
Fig.~\ref{fig:Z_tagging} for TNG300.  At all radii the contribution from SNIa
to the overall metal budget is not very large and metal production is largely
dominated by SNcc. This is expected and also in agreement with observations
that find based on the relative metal abundances in the outskirts of clusters
that the enrichment should have been dominated by SNcc with only a small metal
fraction coming from SNIa~\citep[][]{Simionescu2015}.  Furthermore, the
contribution of AGB stars to the iron budget is negligible and SNIa dominate
the iron production as expected. We also note that the scatter between the
different source fractions is small, indicated by the contours. Especially, in
the outer parts we find a very low dispersion within our cluster sample for the
fractional metal and iron contributions from different sources. There is,
therefore, also a strong uniformity among all $370$ clusters regarding the origin of
metals and iron at most radii within the ICM. The three panels also demonstrate
that fractional source profiles for metal and iron are rather flat in the
outskirts of the clusters, and similar for all redshifts back to $z \sim 2$.
Towards the center the iron contribution from SNIa turns slightly up while at
the same time the contribution from SNcc goes  down. We see similar
trends for the total metal content: in the outer parts of the cluster SNcc
dominate the overall metal budget. However, towards the center the contribution
of AGB stars is increasing. 

\begin{figure*}
\centering
\includegraphics[width=0.495\textwidth]{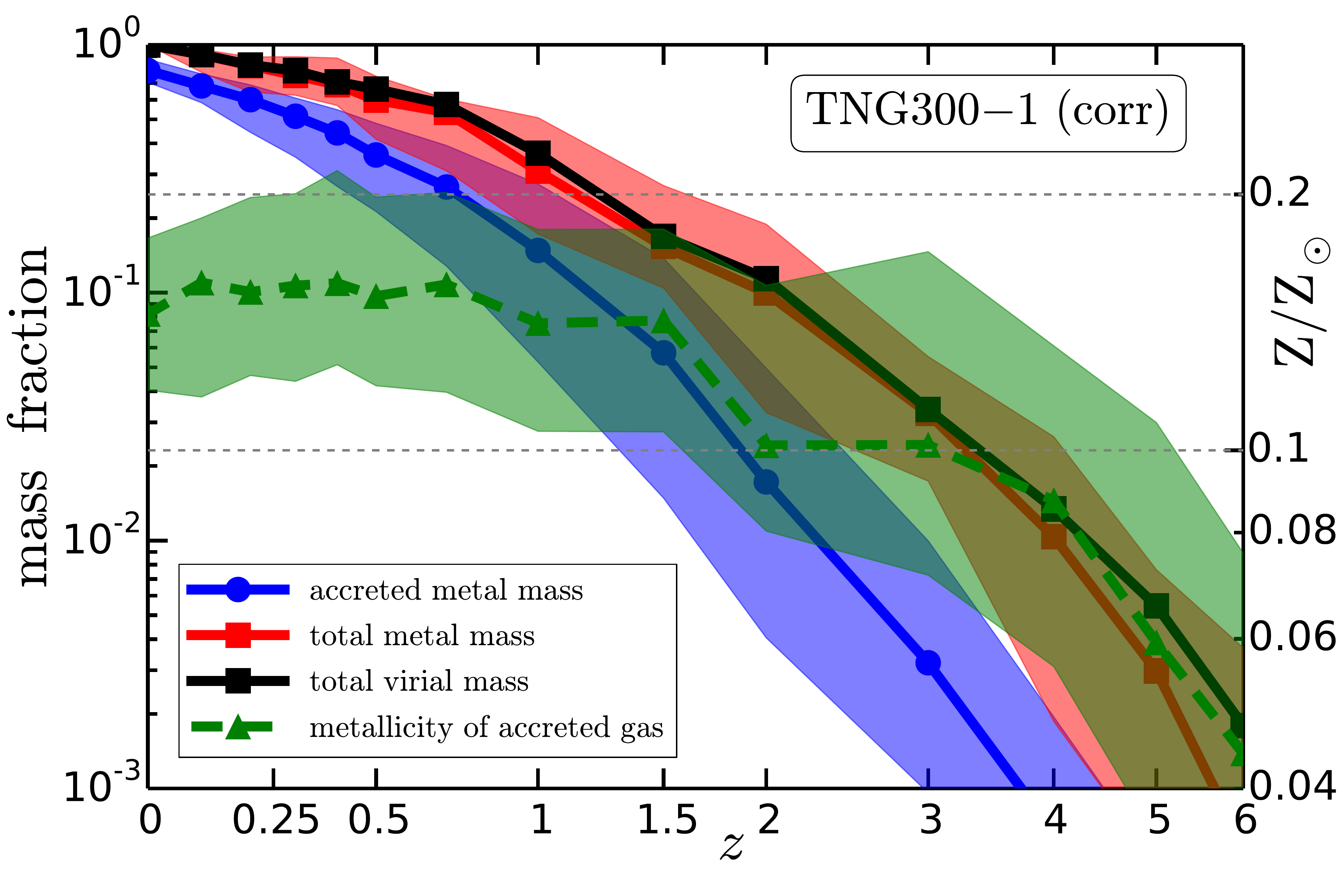}   
\includegraphics[width=0.495\textwidth]{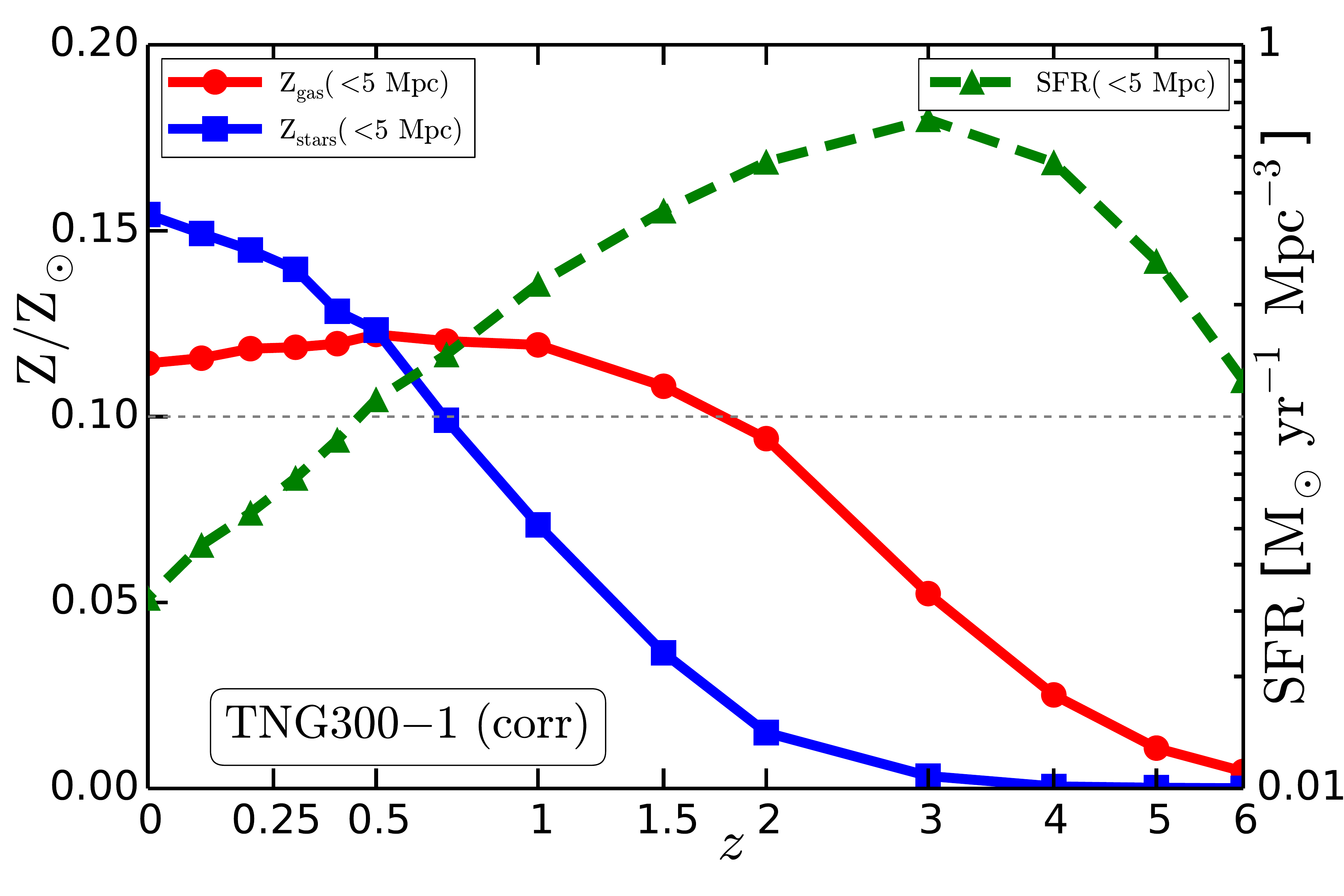}\\
\vspace{0.5cm}
\includegraphics[width=0.3\textwidth]{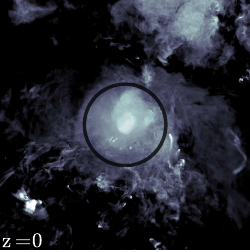}
\includegraphics[width=0.3\textwidth]{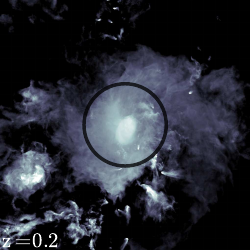}
\includegraphics[width=0.3\textwidth]{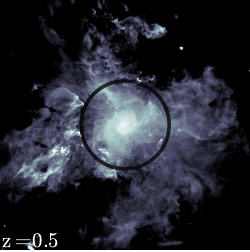}
\includegraphics[width=0.3\textwidth]{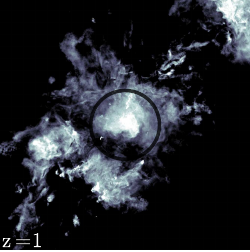}
\includegraphics[width=0.3\textwidth]{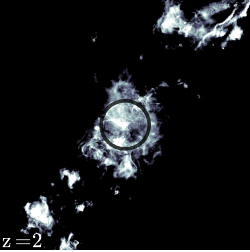}
\includegraphics[width=0.3\textwidth]{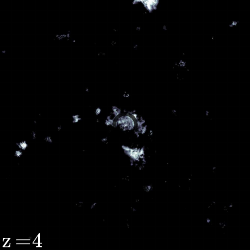}
\includegraphics[width=0.8\textwidth]{figures/maps/colorbar_fof_0_12_z0_Z_TNG300.pdf}
\caption{Upper left panel: Median cumulative accreted metal mass fraction within
$r_{\rm 200, crit}$ as a function of redshift. In total more than $\sim 80\%$
of all metals within the virial radius at $z=0$ have been accreted from the
IGM; i.e. has an ex-situ origin. The shaded region indicate the variation of this fraction within the
cluster sample. This scatter is small towards $z=0$ demonstrating that most
clusters in the sample have accreted a similar amount of metals. We also show
the metal mass within $r_{\rm 200, crit}$ at each redshift, and the average
metallicity of accreted gas. The latter does not evolve much since $z \sim 2$,
i.e. despite the fact that significant metal mass is accreted afterward, this
gas has a rather uniform metallicity which causes the outer metallicity
profiles to be nearly constant as a function of time. The black line shows
the growth of the median total mass of the cluster sample. Upper right panel: Median of
the average metallicities of gas and stars, and star formation rate density as
a function of redshift measured in spheres with a comoving radius of $5\Mpc$
around the cluster progenitors. The proto-cluster environment is enriched at
the time when the star formation rate peaks. This enriched material is then
later accreted onto the clusters. The enrichment of stars is delayed compared
to the gas since those stars have to form from enriched gas first. Lower panels: Metallicity maps for the most massive cluster of TNG300 at different redshifts. The extent of the maps are $10\Mpc$ comoving, i.e. covering
the region over which averaged quantities are measured in the upper right panel. Circles indicate $r_{\rm 500, crit}$.
}
\vspace{-0.5cm}
\label{fig:metal_tracing}
\end{figure*}

\cite{Biffi2017} performed a similar analysis for their zoom-in cluster sample.
However they inspected only a small number of clusters for the origin of metals, so a statistically meaningful
comparison with our findings is not possible. Nevertheless, the ratio of metals
produced from SNcc over metals produced by SNIa decreases towards the cluster
center in our simulations, whereas the ratio of metals from SNcc to metals from
SNIa slightly increases at smaller radii for the cluster sample
in~\cite{Biffi2017}. This directly impacts the radial profiles of the
silicon over iron and oxygen over iron ratios since silicon and oxygen are
produced by SNcc, whereas iron is produced by SNIa. We therefore expect that
those abundance ratios should follow the trend seen for the metal sources in
Fig.~\ref{fig:Z_tagging}.  Indeed, as demonstrated in the previous section, we
find that the oxygen over iron and silicon over iron ratios slightly decrease
towards the cluster centers, which is a consequence of the enrichment pattern
seen in Fig.~\ref{fig:Z_tagging}. On the contrary, \cite{Biffi2017} find
slightly increasing silicon over iron and oxygen over iron ratios towards the
cluster centers in agreement with their different enrichment history reflected
by the different contributions from SNIa and SNcc as a function of radius in
their analysis. In both cases, and as expected, those abundance ratios are good
tracers to differentiate between different enrichment histories. However, it also seems
that the different galaxy formation models and enrichment schemes of the two
simulations lead to different trends for the detailed enrichment history within
the ICM. 

Recent observations favor rather flat abundance ratios for silicon over iron
and oxygen over iron~\citep[e.g.,][]{Mernier2017} as shown in
Fig.~\ref{fig:element_profiles}.  These observational abundance ratios can also be
fit with a combination of SNIa and SNcc yield models to quantify the radial
contribution of SNIa and SNcc products. \cite{Mernier2017} shows that this
leads to quite constant contributions from SNIa and SNcc, which seems to be
slightly inconsistent with the trends seen in current simulations.  We also
include some observationally derived SNIa fraction in the top panel of Fig.~\ref{fig:Z_tagging}
based on~\cite{Ezer2017}. They find for Abell 3112 that the ratio of the number
of SNIa explosions to SNcc has a uniform distribution at a level of $12-16\%$
out to the virial radius. We convert this ratio to an overall metal mass
production ratio to be consistent with the results presented in
Fig.~\ref{fig:Z_tagging}. Our simulations predict a lower SNIa metal
contribution compared to these observations although the outer data points are
marginally consistent with the simulation predictions within the error bars.
In general, this nearly constant SNIa contribution found in observations suggests
that the ICM in cluster outskirts has been enriched by metals at an early stage
before the cluster itself was formed during the period of intense star
formation activity. Overall this finding agrees with our results, although our
SNIa contribution seems to be slightly smaller.

We can also use the metal tagging technique to understand the origin of 
iron in clusters as a function of redshift as presented in
Fig.~\ref{fig:median_Z_redshift}. In Fig.~\ref{fig:median_tag_redshift} we show
the median fraction of iron mass originating from SNIa and SNcc as a function
of time for the cluster sample of TNG300. As expected the SNcc contribution
dominates at early times due to the short SNcc timescales. At later times the
contribution from SNIa dominate since SNIa operate on much longer timescales.
The figure also reveals that we expect an equal contribution from SNIa and SNcc
at around $z\sim 2.5$. At earlier times SNcc dominate the iron budget, whereas
SNIa take over at later times.  The transition is rather quick such that below
$z \sim 2$ the contribution of SNIa versus SNcc is quite constant in agreement
with the profiles presented in Fig.~\ref{fig:Z_tagging}. 

To get more insights into the origin of ICM metals and the reason for the uniformity and time-invariance of the corresponding distributions, we track the
amount of metal mass accreted onto the cluster from beyond its virial radius
using the tracer particles~\citep[][]{Genel2013}.  In the upper left panel of
Fig.~\ref{fig:metal_tracing} we present the cumulative fraction of accreted
metal mass as a function of redshift for the cluster sample of TNG300. We find
that more than $\sim 80\%$ of all metals within $r_{\rm 200, crit}$ at $z=0$
have been accreted during the evolution of the cluster sample.  This number is
quite universal and robust across the cluster sample as can be seen by the
small scatter of the accretion fraction at low redshifts. We note that we only
count metals entering the virial radius in the form of gas; i.e. metal enriched
gas returned by infalling stars is not accounted for. Therefore, the $\sim
80\%$ value should be interpreted as a lower limit. We also include in the upper left 
panel of Fig.~\ref{fig:metal_tracing} the normalised total mass of metals
within $r_{\rm 200, crit}$ at each redshift. At $z\sim 2$ the cluster has
assembled only about $10\%$ of its final metal mass, which is consistent with
the fact that at that time only a small fraction of metals have been accreted. The black line shows
the growth of the median total mass of the cluster sample. 

Although these results seem to be consistent with each other they seem to
contradict the results we have presented in
Fig.~\ref{fig:metal_profiles_redshift}, where we have demonstrated that the
outer metallicity profiles are time-invariant since around $z\sim 2$. This
inconsistency can be understood by looking at the average metallicity of the
gas that is accreted onto the clusters, which we also show in in the upper left panel
of Fig.~\ref{fig:metal_tracing} (dashed line). Below $z \sim 2$ the
average metallicity of accreted gas is nearly constant and close to ${\rm
Z}/{\rm Z}_\odot \sim 0.1$, which is consistent with the constant metallicity
profiles in the outer cluster regions discussed above. The accreted gas has
therefore been enriched before accretion, so that the average metallicity
within the cluster stays rather constant in the outer parts in agreement with
the findings presented above. We note that the metallicity of the accreted gas varies only between ${0.05\,{\rm Z}_\odot \lesssim {\rm Z} \lesssim 0.2\,{\rm Z}_\odot}$ since $z \sim 2$ over the cluster sample. Therefore, the metallicity of accreted gas is also quite universal for our cluster sample.

In the upper right panel of Fig.~\ref{fig:metal_tracing} we quantify the early
enrichment of the proto-cluster environment in more detail.  Here we present
the median mass-averaged gas and stellar metallicity within a comoving radius
of $5\Mpc$ around the cluster progenitors at various redshifts along with the
comoving star formation rate density in the volume. We exclude the gas within
$r_{\rm 200, crit}$ at each redshift. The median star formation rate peaks
around  $z \sim 3$ and leads to significant enrichment of the nearby gas below
$z \sim 2$. This median metallicity is consistent with the accreted metallicity
seen in the upper left panel. At the same time the average stellar metallicity is
also increasing towards lower redshifts. However, this occurs only after some
delay since the stars have to form from the enriched gas first. In the lower panels we present metallicity maps for the most massive halo of TNG300 at different redshifts. The extent of the maps is $10\Mpc$ comoving, i.e. covering
the region over which averaged quantities are measured in the upper right panel. Those maps also demonstrate the high average
metallicity in the proto-cluster environment already at early times. This enriched material is then accreted towards lower redshift and is the reason
for the time-invariant metallicity profiles in the outer parts of the clusters.

We conclude that our galaxy formation model is able to reproduce broadly also
the high redshift metallicity measurements. Combined with the low-redshift
results presented in the previous section, we arrive at a self-consistent
enrichment history which seems to favor an early enrichment. However, we also
note that the detailed contribution of SNIa and SNcc to the metal budget and
consequently the abundance ratios are slightly inconsistent with most recent
observational data. A natural extension of the analysis presented here would be a study to
investigate the exact origin of the metals within the proto-cluster environment
using the tracer particles.

\section{Conclusions}\label{sec:conclusions}

We have presented a first analysis of the metal content of galaxy clusters of the
new IllustrisTNG simulations~\citep[][]{Springel2017, Marinacci2017, Naiman2017, Pillepich2017a, Nelson2017}. IllustrisTNG is the follow-up project of
Illustris~\citep{Vogelsberger2014, Illustris, Genel2014, Sijacki2015}, and
consists of three different main simulations TNG50, TNG100, TNG300 each at three
different resolution levels. The side length of {\bf TNG100} ($2 \times 1820^3$ resolution elements) is $110.72\Mpc$ with a
mass resolution of $7.46\times 10^6\msun$ and $1.39\times 10^6\msun$ for the DM
and baryonic components, respectively. {\bf TNG300} ($2 \times 2500^3$ resolution elements) has a side length of
$302.63\,\Mpc$ and a mass resolution of $5.88\times 10^7\msun$ and $1.1\times
10^7\msun$ for the DM and baryonic components, respectively. {\bf TNG50} ($2 \times 2160^3$ resolution elements) has a side
length of $51.67\,\Mpc$ and a mass resolution more than an order of magnitude
higher than TNG100.

Here we have focused on the analysis of two cluster
samples ($M_{\rm 500, crit} > 10^{13.75}\msun$) of the TNG100 ($20$ clusters)
and TNG300 ($370$ clusters) simulations, which contain sizeable cluster populations.  
We find that the low-redshift metallicity profiles in those simulations agree
with X-ray inferred observed abundance profiles. Furthermore, the simulations also
reproduce high-redshift evolutionary trends of the ICM metal content. Our main
findings are:

\begin{itemize}
\item The radial median metallicity profiles agree with observational X-ray
data, and the clusters in the different simulations show similar profiles
across a wide mass range. The median metallicity, iron, silicon and oxygen
profiles follow shallow negative gradients with very little scatter in the
outskirts of the ICM ($\sigma_{\rm Z}/{\rm Z} \sim 15\%$ relative scatter). This scatter increases to
$\sigma_{\rm Z}/{\rm Z} \sim 40\%$ towards the centers of the clusters ($\sim 0.01 \times r_{\rm
500, crit}$). The central metallicity reaches $\sim 0.4 - 1.0\,{\rm Z}_\odot$ ,
and drops to $\sim 0.2 \,{\rm Z}_\odot$ at larger radii ($\sim r_{\rm 500,
crit}$).  The central logarithmic slope of the metallicity profile is close to
$-0.1$ at one percent of $r_{\rm 500,crit}$ and $-0.5$ at $r_{\rm 500,crit}$.
The azimuthal iron scatter around $r_{\rm 500,crit}$ has roughly the same amplitude as
the iron scatter found through Suzaku observations of the Perseus
cluster~\citep[][]{Werner2013}. We note that we have focused here only on total metallicity, iron,
silicon, and oxygen abundance profiles and distributions. However, our galaxy model also traces other elements beyond those. We present
profiles for the other abundance ratios in Fig.~\ref{fig:all_elements}. All profiles are rather flat with a slight positive gradient. The highest ratio
is predicted for neon, whereas nitrogen has the lowest values at all radii.

\vspace{0.3cm}

\item Cool core clusters in our simulations, classified by their central
being less than $30\,{\rm keV}\,{\rm cm}^{-2}$~\citep[][]{Hudson2010},  exhibit radial metallicity
distributions, which are steeper and peaked towards the cluster centers
compared to those of non-cool core clusters in agreement with observational
data~\citep[][]{Ettori2015}. Cool core clusters approach $0.8\,{\rm Z}_\odot$
towards their centers whereas non-cool core clusters have central metallicities
below $0.5\,{\rm Z}_\odot$. The metallicity profiles of both overlap beyond
$0.15 \times r_{\rm 500, crit}$ in agreement with observational
data~\citep[][]{Ettori2015}.

\vspace{0.3cm}

\item Silicon over iron and oxygen over iron ratios show small positive
gradients, and are therefore in slight tension with most recent observations.
The ratios increase by about $50\%$ from $10^{-2} \times r_{\rm 500,crit}$ to $r_{\rm
500, crit}$. The increasing ratios of these two elements are
consistent with the radial distribution of the production sites of metals in
our simulations. Towards the center the relative contribution of SNIa compared
to SNcc is slightly rising resulting in a decrease of the X/Fe ratios towards
the centers.  

\vspace{0.3cm}

\item The distribution of the mass-weighted averaged metallicities within
$r_{\rm 500, crit}$ can be described by narrow Gaussians with a dispersion of
$0.027\,{\rm Z}_\odot$. We find a small anti-correlation for the temperature-iron abundance
relation with a logarithmic slope, $\sim -0.26$, in agreement with the compiled analysis of~\cite{Yates2017}.
However, the overall normalisation of the temperature-iron abundance relation
is about $0.2\,{\rm dex}$ too low in the simulation compared to observational data.

\vspace{0.3cm}

\item The outskirts, $r \sim r_{\rm 500, crit}$, of the metallicity, iron,
silicon and oxygen profiles do not evolve since redshift $\sim 2$. Therefore, the outer cluster abundance profiles beyond $0.4 \times r_{\rm
500,crit}$ are constant over time with nearly no evolution over the
last $\sim 10\Gyr$. The inner
profiles, $r < 0.1 \times r_{\rm 500, crit}$, are flattening towards lower
redshift and decrease in their amplitude. The central amplitudes decrease by
nearly a factor of $2$ at the cluster centers from $z=2$ to $z=0$. This effect is not seen in observations~\citep[e.g.,][]{Mantz2017}.
It is unclear whether this points to a systematic uncertainty in the observations,
or whether the central enrichment in simulations, and related to this the AGN feedback, are inconsistent with the observed behaviour at cluster centers.

\begin{figure}
\centering
\includegraphics[width=0.495\textwidth]{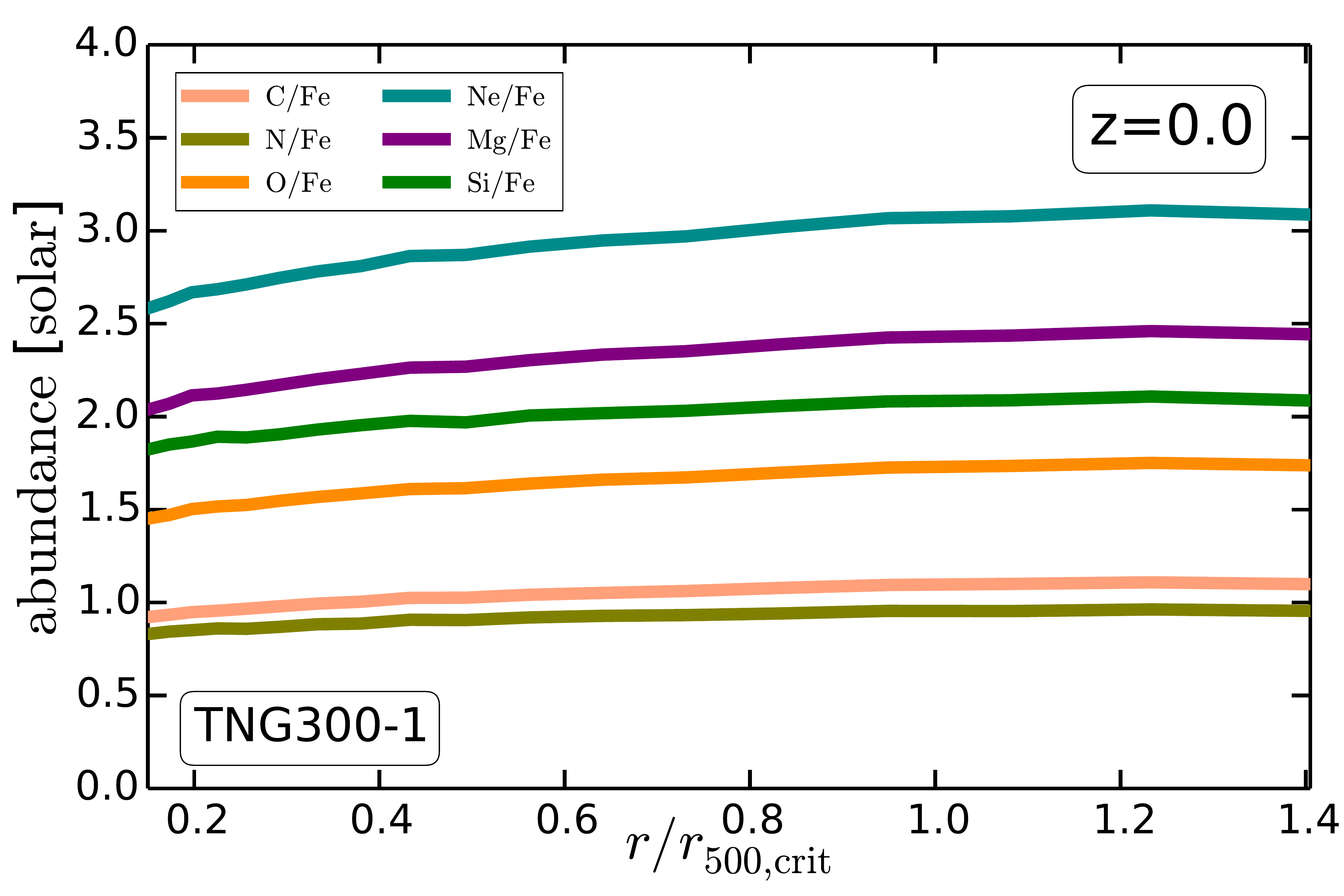}
\caption{Median abundance ratio profiles of carbon, neon, nitrogen, magnesium, oxygen, and silicon over
iron. We show for each ratio the median profile of TNG300-1. These are all
non-primordial elements tracked by our galaxy formation model.}
\vspace{-0.5cm}
\label{fig:all_elements}
\end{figure}

\vspace{0.3cm}

\item The averaged metallicities within the ICM ($r<r_{\rm 500, crit}$) do not
evolve with redshift since $z \sim 1$ in agreement with recent observational
findings of~\cite{McDonald2016}.  Based on our simulations we find that the
true physical scatter of iron abundances at various redshifts is significantly
smaller ($\sim 0.03\,{\rm Fe}_\odot$) than the observational variation, which is dominated by measurement 
uncertainties.  The iron is produced at early times by SNcc (beyond $z\sim 2.5$) and at
late times through SNIa below $z \sim 2.5$.  The transition of the
two regimes is rather quick such that the contributions of SNcc and SNIa do not
evolve significantly below $z\sim 2$. The average metallicity within $0.1
\times r_{\rm 500, crit}$ is evolving from $z=0$ to $z=1$ with an increase of
about $\sim 20\%$ towards $z \sim 1$.  We compare the
evolution of this central average iron abundance to the best-fit model of~\cite{Mantz2017}, and find that
the average enrichment level agrees with the simulation data, but we see a different
redshift evolution. 

\vspace{0.3cm}

\item We find no central metallicity drops~\citep[][]{Mernier2017} in the
clusters of the IllustrisTNG simulations. However, we observe those for
clusters of the Illustris simulation where a different radio mode AGN model has
been employed. This difference might point towards bubble-driven uplifting of
metals through AGN feedback in Illustris, but not in IllustrisTNG since the
latter simulation suite uses a kinetic AGN feedback model at the cluster center
not supporting such a mechanism at the same strength. For Illustris the median metallicity profiles
peak at $0.03\times r_{\rm 500, crit}$ reaching $\sim 0.7\,{\rm Z}_\odot$ and
drop to $\sim 0.6\,{\rm Z}_\odot$ at $10^{-2}\times r_{\rm 500, crit}$. This
is comparable to the central oxygen drop seen in~\cite{Mernier2017}, but larger
than the observer iron drop therein.

\vspace{0.3cm}

\item At least $\sim 80\%$ of the metals within the virial radius ($r_{\rm
200,crit}$) are accreted. This fraction is universal across the cluster sample
with very little scatter. The accreted gas is enriched to $\sim 0.1\,{\rm
Z}_\odot$ at around $z \sim 2$ slightly after the peak of star formation in the
proto-cluster environment. At $z\sim 2$ the cluster has assembled only about
$10\%$ of its final metal mass, which is consistent with the fact that at that
time only a small fraction of metals have been accreted yet. The proto-cluster
environment for the clusters is already enriched to $> 0.1\,{\rm Z}_\odot$
below $z \sim 2$ briefly after the peak of the local star formation. This
proto-cluster gas is then accreted onto the cluster and leads to the constancy
of the outer metallicity profiles from $z \sim 2$ to $z = 0$.

\vspace{0.3cm}

\item The detailed shapes of the metallicity profiles and metal
distributions are converged. The normalisation however depends on the degree of
convergence of stellar mass. Any relative ratios derived from the simulation,
e.g., abundance ratios profiles, are therefore converged. Furthermore, low
resolution results can be extrapolated to higher resolutions through a scaling
with the stellar mass ratio of the two resolutions. Specifically, the metal
content of clusters of TNG300-1 can be scaled to TNG100-1 resolution by
applying a correction factor of $1.6$ which follows from the stellar mass ratio
between the two simulations.  

\end{itemize}

We conclude that the predictions of IllustrisTNG
are generally in good agreement with observational findings at low and high
redshifts. 
Overall we find that the abundance profiles of clusters and the 
enrichment exhibit a strikingly uniform behaviour over our large cluster
sample. This is true for the radial metallicity profiles, and also for the
metal accretion history. Besides this uniformity, cluster average metallicities
and metal profiles are also remarkably constant as a function of time. The TNG300
cluster sample with $370$ objects demonstrate these characteristics in a
convincing way through the large sample size, which is currently unparalleled.

Further progress in understanding the origin and evolution of metals in
clusters has to come from two fronts: observationally and theoretically.
Observationally, \cite{McDonald2016} and \cite{Mantz2017} presented the so far most detailed
accounting of the enrichment history of the ICM. A more refined analysis will
require next-generation observatories, such as XARM, Athena and X-ray Surveyer,
combined with samples of clusters at $z \sim 2$ which will be available with
the next generation of Sunyaev-Zel'dovich experiments. Significant progress in measuring
metallicity profile will also come with next generation X-ray observatories.
The larger effective area of future missions will also allow studies of the
outskirts of clusters in more detail. This combined with increased spectral
resolution will lead to new insights into the distribution of metals in the ICM
of clusters.  From the theory side there still seems to be no perfect consensus
on reproducing all details of observed abundance trends among different simulations. The detailed metal
distribution, especially at cluster centers, also depends sensitively on
details of the galaxy formation model, like the AGN feedback implementation, as
we have demonstrated.  Ultimately, more detailed galaxy formation models should
be employed going beyond the coarse sub-resolution models used in simulations like
IllustrisTNG. Those models should include more refined schemes for stellar
winds and AGN feedback.  Both of them affect the gas enrichment and should
therefore be modelled more faithfully at resolutions beyond those of
IllustrisTNG. In fact, even at the TNG100-1 resolution more
detailed AGN feedback models are applicable~\citep[e.g.,][]{Weinberger2017b},
which might lead to differences in the central metal distribution. Higher
resolution cluster simulations should therefore definitely explore those new
models and contrast the results with those of coarser models like those of
Illustris and IllustrisTNG.  Last, also other physical processes like thermal
conduction~\citep[][]{Kannan2017} or dust formation~\citep[][Vogelsberger et
al., in prep]{McKinnon2016, McKinnon2017} affect the metallicity profiles in
and around galaxies and should therefore be considered in future models.

\section*{Acknowledgements}
We thank the anonymous referee for many useful comments that helped to
improve the presentation of our results. We are grateful to VRG for sharing
the codes to generate the merger trees of the new simulations. MV thanks David
Barnes, Irina Zhuravleva, Michael McDonald, and Esra Bulbul for useful
discussions.  VS, RW, and RP acknowledge support through the European Research
Council under ERC-StG grant EXAGAL-308037 and would like to thank the Klaus
Tschira Foundation. MV acknowledges support through an MIT RSC award, the
support of the Alfred P. Sloan Foundation, and support by NASA ATP grant
NNX17AG29G. The Flatiron Institute is supported by the Simons Foundation. SG
and PT acknowledge support from NASA through Hubble Fellowship grants
HST-HF2-51341.001-A and HST-HF2-51384.001-A, respectively, awarded by the
STScI, which is operated by the Association of Universities for Research in
Astronomy, Inc., for NASA, under contract NAS5-26555. JPN acknowledges support
of NSF AARF award AST-1402480.  The flagship simulations of the IllustrisTNG
project used in this work have been run on the HazelHen Cray XC40-system at the
High Performance Computing Center Stuttgart as part of project GCS- ILLU of the
Gauss Centre for Supercomputing (GCS).  Ancillary and test runs of the project
were also run on the Stampede supercomputer at TACC/XSEDE (allocation
AST140063), at the Hydra and Draco supercomputers at the Max Planck Computing
and Data Facility, and on the MIT/Harvard computing facilities supported by FAS
and MIT MKI.

\label{lastpage}

\end{document}